\documentclass[ ]{aa}

\usepackage{graphicx}
\usepackage{color}
\usepackage{nicematrix}
\usepackage[varg]{txfonts}
\usepackage{multirow}
\usepackage{makecell}
\usepackage{amsmath}
\usepackage{epstopdf}
\usepackage{gensymb}
\usepackage{lscape}
\usepackage{threeparttable}
\usepackage[round]{natbib}
\graphicspath{{Pictures}}
\title{Wide-band fluctuations of solar active regions probed with SHARP magnetograms}

\author{G.~Dumbadze\inst{1,2,3}, B.M.~Shergelashvili\inst{2,3,4}, M.L.~Khodachenko\inst{5} and S.~Poedts\inst{1,6} }

\institute{Centre for mathematical Plasma Astrophysics, Department of Mathematics, KU Leuven, Celestijnenlaan 200B, B-3001, Leuven, Belgium\\
                 \and
         Centre for Computational Helio Studies, Ilia State University, G.\ Tsereteli street 3, 0162 Tbilisi, Georgia\\
                  \and
         Evgeni Kharadze Georgian National Astrophysical Observatory, M.\ Kostava street 47/57, 0179 Tbilisi, Georgia\\
                 \and
         Ruhr-Universität Bochum, Institut für Theoretische Physik IV, 44780 Bochum, Germany\\
                 \and
        Space Research Institute, Austrian Academy of Sciences, Schmiedlstra{\ss}e 6, 8042 Graz, Austria\\
                  \and
        Institute of Physics, University of Maria Curie-Sk{\l}odowska, Pl.\ M.\ Curie-Sk{\l}odowska 1, PL-20-031 Lublin, Poland
}

\begin{document}

\abstract
{The power spectra of the fluctuation noise of the solar active region (AR) areas and magnetic fluxes sequentially observed in time contain information about their geometrical features and the related fundamental physical processes. These spectra are analysed for five different ARs with various magnetic field structures. }
{The goal of this work is to detect the characteristic properties of the Fourier and wavelet spectra evaluated for the time series of the fluctuating areas and radial magnetic fluxes of the active regions.  Accordingly, this work gathers information on the properties of noise in the different cases considered. }
{The AR area and radial magnetic flux time series were built using SHARP magnetogram datasets that cover nearly the entire time of the ARs' transits over the solar disk. Then we applied Fourier and wavelet analyses  to these time series using apodization and detrendization methods for the cross-comparison of the results. These methods allow for the detection and removal of the artefact data edge effects. Finally, we used a linear least-squares fitting method for the obtained spectra on a logarithmic scale to evaluate the power-law slopes of the fluctuation spectral power versus frequency (if any). }
{According to our results, the fluctuation spectra of the areas and radial magnetic fluxes of the considered ARs  differ from each other to a certain extent, both in terms of the values of the spectral power-law exponents and their frequency bands. }
{The characteristic properties of the fluctuation spectra 
for the compact, dispersed, and mixed-type ARs exhibit noticeable discrepancies amongst each other. It is plausible to conclude that this difference might be related to distinct physical mechanisms responsible for the vibrations of the AR areas and/or radial magnetic fluxes. }


\titlerunning{Wide-band spectra of solar active regions from SHARP datasets}

\authorrunning{Dumbadze, Shergelashvili et al.}

\maketitle

\section{Introduction}
The solar interior and its atmosphere consist of a complex system of magnetically shaped, dynamic or static plasma features. The topology of the magnetic field carries information on the formation of the cyclically variable large-scale solar magnetic field and its fragmentation into smaller structures related to sunspots, magnetic loops, filaments, prominences, and so on. On the other hand, these structures are able to sustain various periodic and aperiodic, as well as background stochastic processes, of the plasma and magnetic field motion or modification. Their random nature in many cases indicates the presence of substantially non-equilibrium random driving mechanisms \citep[e.g.~see][]{Maes2009}.  

In our previous studies of solar active regions (ARs), their quasi-periodic behaviour has been reported in \citet[][referred to as Paper~I throughout the text]{Dumbadze17} and in \citet[][referred to as Paper~II throughout the text]{Dumbadze21}, where different types of AR oscillations with the characteristic periods of 9-10~h, 6-7~h, and 3.5-4~h have been revealed using magnetogram
datasets of Solar Dynamics Observatory \citep[SDO;][]{Pesnell2012} Helioseismic and Magnetic Imager instrument \citep[HMI;][]{Scherrer12,Schou12}.

These long-period processes seem to be permanently present in the ARs as some kind of 'undamped' oscillations on top of visibly manifested background noise. 
Because of the relatively low signal-to-noise ratios (S/Ns), special techniques have been applied to reveal the characteristics of these oscillations. This behaviour is similar to the recently observationally detected short-period undamped transverse oscillations in  coronal loops \citep{Nakariakov2022,Nakariakov2021,Nakariakov2016}. Although the actual character (e.g. transverse or longitudinal) of the long-period coherent fluctuations described in Paper~I and Paper~II is not yet clear, the similarity of the dynamic pictures with those taking place for the short-period transverse oscillations in the coronal loops persists. In addition, the existence of a myriad of small-scale, stochastically distributed aperiodic motions of EUV intensity enhancements were  recently reported  \citep{Shergelashvili2022}. These were detected and categorised using the statistical categorisation method developed in \citet{Philishvili2021}. It is plausible to believe that the aforementioned long-period AR oscillations are also strongly dependent on the background noise. Therefore, the investigation of the noise fluctuation spectra accompanying the earlier discovered coherent signals is the main subject of the work presented here.

\citet{Auchere2014} analysed the intensity data cube, $195\;$\AA, provided by the Solar and Heliospheric Observatory \citep[SOHO;][]{Domingo1995}, with the Extreme ultraviolet Imaging Telescope \citep[EIT;][]{Delaboudiniene1995}, and identified 917 events of  solar origin. The intensity pulsations with periods of 3 to 16~h took place in these events on top of an approximately power-law distributed background noise spectrum. However, \citet{Auchere2014} could not identify the physical cause of these pulsations. 
\citet{Froment2015} repeated the analysis of \citet{Auchere2014} for the Atmospheric Imaging Assembly \citep[AIA;][]{Lemen2012} instrument of SDO and found more than 2000 intensity pulsation events across a similar long-period range. For these events, they specified  three dominant power peaks at around 9.0~h, 5.6~h, and 3.8~h, respectively, which were also addressed in \citet{Auchere2016}. In addition, ultra-long-period oscillations ($\sim10-30\;$h) were found in the EUV filaments by \citet{Foullon2004} and \citet{foullon09}, using the SOHO/EIT $195\;$\AA~datasets. 

\citet{Kolotkov2017} used the line-of-sight (LOS) SDO/HMI magnetograms to investigate 13 hours of observational data on small-scale magnetic flux structures. They performed a spectral analysis, employing the Hilbert-Huang transform (HHT) technique that decomposed the signal into 11 empirical modes, where the last one was "a long-term aperiodic trend of the signal." According to \citet{Kolotkov2017}, the remaining nine empirical modes were related to the noisy components and one mode was shown to have a gradually increasing oscillation amplitude and its oscillation period increased with amplitude from $\sim80$ to $230\;$min." \citet{Efremov2014,Efremov2018} analysed the dynamics of the magnetic field strength in sunspot SOHO/MDI magnetograms. They suggested that the detected long-period (periods 10-35~h) oscillations of the magnetic field in sunspots are related to the oscillations of the sunspots as a whole.

All the above-mentioned significant period peaks are usually embedded in a strong background noise that is observable in the power spectra of various frequency bandwidths.
\citet{Leonardis2012} used the Hinode Solar Optical Telescope (SOT) to investigate the power spectral densities of five areas with  quiescent prominences on the northwest solar limb. They obtained power-law spectra, $W \propto f^{-|m|}$, with an approximate exponent value $|m| \approx 1.2$.  \citet{Gupta2014} obtained almost the same values, namely, of $|m| = 1.2-1.7$, where SDO/AIA $171\;$\AA~and 193$\;$\AA~images were used and the Fourier power spectra of the intensity fluctuations at six locations of the south polar coronal hole were analysed.
\citet{Ireland2015} used SDO/AIA $171\;$\AA~and $193\;$\AA~to analysed six hours of image data at four locations of the image field: a quiet sun region, a region that lies on top of a sunspot, a footpoint region, and a coronal moss region. The time series of the images revealed approximately power-law properties with values of $|m|$ in the range of $1.8-2.3$. \citet{Kolotkov2016} studied the statistical properties of EUV emission above the centre of a sunspot and above a quiet sun region, using SDO/AIA $304\;$\AA, $171\;$\AA~and $1600\;$\AA~data. In both regions, values of $|m|$ in the range of $0.86-1.33$  were found in the obtained power-law spectra.

In this paper, we study the power spectra of the background noise using the SDO/HMI datasets obtained during the transit of ARs over the visible solar disk. We consider five different ARs and analyse the properties of their power-law Fourier spectra. The remainder of this paper is organised as follows. The observations and data processing are described in Sect.~\ref{secdata}. The results and discussions are given in Sect.~3. Finally, the conclusions are presented in Sect.~4.

\section{Observations and data processing}\label{secdata}

\begin{table}[!ht]
\caption{Selected ARs. The values of longitudes and latitudes indicate the AR location on the solar surface at the beginning and the end moment of the observation period. While the area values are taken at the moment when the centre of mass of the AR is at or very close to the visible central meridian. }
 \centering
\scriptsize
\begin{threeparttable}
\begin{tabular}{cccrrc}
 \hline \hline
AR & Obs. time & \multicolumn{1}{c}{Longitude}    & \multicolumn{1}{c}{Latitude}    & \multicolumn{1}{c}{Area\tnote{2}}    \\
                no.   & (Hours)   & \multicolumn{1}{c}{($\degree$)}  & \multicolumn{1}{c}{($\degree$)} & \multicolumn{1}{c}{(km$^2$)}   \\
 \hline
12381 & $240$ & [$-69.9$ \; $65.2$] & [$14.7$  \; $14.5$]  & $1.6\cdot10^9$  \\
12378 & $240$ & [$-64.5$ \; $60.2$] & [$-15.7$ \; $-14.9$] & $1.5\cdot10^8$  \\
12435 & $225$ & [$-62.5$ \; $54.4$] & [$-15.3$ \; $-15.4$] & $6.0\cdot10^7$  \\
12437 & $225$ & [$-57.9$ \; $59.2$] & [$-21.6$ \; $-21.7$] & $6.0\cdot10^7$  \\
12524 & $240$ & [$-69.2$ \; $69.6$] & [$14.9$  \; $14.3$]  & $6.1\cdot10^8$  \\
 \hline
\end{tabular} \label{tableAR}
    \end{threeparttable}
\end{table}

Time series of the magnetic field radial component $B_r$, observed by SDO/HMI, have been taken from the Space-weather HMI Active Region Patches \citep[SHARP;][]{Bobra14} database.
They include rectangular frames that enclose the corresponding AR and track it along the transit over the visible solar disk. 
The tracked AR patches are remapped to a Lambert Cylindrical Equal-Area (CEA) projection.
We considered the same set of five active regions investigated in our previous study in the Paper-II during their transit over the visible solar disk (see Table~\ref{tableAR} for parameters of the selected ARs).  All the observational features of the considered ARs, which are unimportant for the present study, can be found in the above mentioned Paper~II. We here build time series of characteristic physical quantities of the ARs using the same methodology as in Paper~II. To investigate the power spectra elaborated from the data time series for the total cross sectional areas and the corresponding mean and total magnetic fluxes (while distinguishing between the signed and unsigned, positive and negative magnetic fields), we perform a Fourier analysis of each dataset, separately. The particular evaluation methods of the characteristic physical quantities are outlined in the following subsections. For clarity and convenience, the acronyms and symbols used are summarised in Table~\ref{tablex}.

\begin{table*}[!ht]
\caption{Summary of the acronyms and symbols used. }
 \centering
\begin{tabular}{lll}
 \hline  
acronym & meaning & symbol/def.    \\
 \hline
CEA                 &cylindrical equal-area mapping       &                 \\
            &magnetic field threshold             &$th$             \\
            &frequency                            &$f$              \\
psd                 &power spectral densities             &$W$              \\
MNMF            &mean negative magnetic fluxes            &$B_{r-}/N_{th}$  \\
MPMF            &mean positive magnetic fluxes            &$B_{r+}/N_{th}$  \\
MSMF            &mean signed magnetic flux                    &$B_r/N_{th}$     \\
MUMF            &mean unsigned magnetic flux              &$|B_r|/N_{th}$   \\
            &number of selected active pixels with $|B_r|>\;th$  &$N_{th}$  \\
            &size of pixel on solar surface       &$S_{CEA}$        \\[-4pt]
            &{\em (set by the CEA projection mapping)}  &           \\
SE              &standard error                           &                 \\
TNMF            &total negative magnetic fluxes           &$B_{r-}$         \\
TPMF            &total positive magnetic fluxes           &$B_{r+}$         \\
TSMF            &total signed magnetic flux                   &$B_r$            \\  
TUMF            &total unsigned magnetic flux             &$|B_r|$          \\
            &areas of concentration of magnetic field above threshold  &$\Sigma_{th}$  \\
            &radial magnetic flux in the areas of magnetic field concentration  &$\Phi_{th}^{I}$  \\
AVSS            &absolute value of the spectrum slope & $\left |m_{\Sigma_{th}^J}\right |$, $\left |m_{\Phi_{th}^{I}}\right |$   \\
            &range of AVSS for areas                       &$\left [ \left |m_{\Sigma_{th}^J}\right | \right ]$ \\
            &range of AVSS for radial magnetic fluxes      &$\left [ \left |m_{\Phi_{th}^{I}}\right | \right ]$ \\
 \hline
\end{tabular} \label{tablex}
\end{table*}

\subsection{The power spectra}
Before carrying out a Fourier spectral analysis of the obtained time series, we performed the standard pre-processing of the data to remove the first, second, or third-order polynomial trends, where relevant. In addition, we applied a normalisation procedure in the datasets as $(X - \overline{X}) / \sigma_X $, where $X$ is the data count from the corresponding dataset, whereas $\overline{X}$ and $\sigma_X$ are the average value and standard deviation of $X$, respectively.

To obtain the power spectra, we applied the following procedures. First, we compute the Fourier spectra to get the power spectral densities (psd), $W,$ out of the normalised datasets. Then, we omit the power of the Nyquist frequency, as suggested by \citet{Vaughan2005}. The power $W$ versus frequency $f$ is plotted in the log-space in the figures throughout the text below. Finally, we make a linear model fit of the obtained power spectra $W(f)$ in the log-scale as: 
\begin{equation}
\label{modelline}
\lg W_{model} = m \lg f + b,
\end{equation}
and we determine the coefficients $m$ and $b$ by a least-square fitting procedure using single-modal linear functions, covering the entire frequency domain. 

To find which part of the dataset satisfies the $H_0$ hypothesis, which implies that the value of the fitted line slope is 0 with respect to a chosen slope, $m$, we calculate the $t-$ statistics $t_{m}$ for the slope as follows:
\begin{equation}
t_{m} = \frac{m}{SE(m)}, 
\end{equation}
where 
\begin{equation}
SE(m) = \frac{\sigma}{\sqrt{ \sum_{i=1}^n \left ( \lg f_i - \overline{\lg f} \right ) ^2}}, 
\end{equation}
\begin{equation}
\overline{\lg f} = \frac{ \sum_{i=1}^n  \lg f_i}{n},
\end{equation}
\begin{equation}
\sigma^2 = \frac{ \sum_{i=1}^n \left ( \lg W_i- \lg W_{model,i} \right )^2}{n-2}, 
\end{equation}
here, $SE(m)$ is the standard error of the slope $m$, $n$ is the total number of considered frequency samples, $f_i$, hence, $n-2$ is the corresponding number of degrees of freedom as we elaborate two parameter linear regression. Similarly, for the intercept, we formulate the $H_0$ hypothesis that its value is $b$ and write the corresponding $t-$ statistics $t_{b}$: 
\begin{equation}
t_{b} = \frac{b}{SE(b)} , 
\end{equation}
where 
\begin{equation}
SE(b) = \sigma \sqrt {\frac{1}{n} + \frac{\overline{\lg f}^2}{ \sum_{i=1}^n \left ( \lg f_i - \overline{\lg f} \right ) ^2}},
\end{equation}
is the standard error of the intercept $b$. 

Furthermore, we compare the obtained values of the $t$ statistics with the corresponding table values of the Student's $t$-distribution $t_{(n-2,\alpha /2)}$ for a specific occasion, where $\alpha$ is the size of area under the right tail of the distribution. We used the values from the $t$-distribution tables for comparison with the values obtained by $t_m$.
If 
\begin{equation}
|t_{m}|<t_{(n-2,\alpha /2)},
\end{equation}
it means that $H_0$ hypothesis is approved and the slope is vanishing ($m=0$) with the significance level $\alpha$. In this way, we find the extreme high and low frequency bands of the power spectrum, satisfying the $H_0$ hypothesis.  
The hypothesis is rejected if 
\begin{equation}
|t_{m}| \geq t_{(n-2,\alpha /2)},
\end{equation}
which is equivalent to the requirement that the statistic's $p$-value (i.e. the probability for a random variable to have the Student $t$-distribution with $n-2$ degrees of freedom is larger than $\left|t_{m}\right|$) satisfies
\begin{equation}
p(|t_m|) \leq \alpha.
\end{equation}
This is the criterion for the rejection of the null hypotheses, meaning $m\neq0$. Therefore, the non-zero $m$ value can be evaluated as the value obtained by the least-square fitting with the uncertainties written as follows:
\begin{equation}
m \pm SE(m) t_{(n-2,\alpha/2)}.
\end{equation}
A similar $t$-test was carried out for the intercept as well to obtain its values and error estimations:
\begin{equation}
b \pm SE(b) t_{(n-2,\alpha/2)}.
\end{equation}
Hence, we estimated the slope $m$ and the intercept $b$ for all datasets. 

For the AR datasets (areas and fluxes of ARs with the thresholds $\pm 350\;$G, $\pm 700\;$G and $\pm 900\;$G), we also determined the residuals of the computed power from the model line Eq.~(\ref{modelline}) assuming that they follow a second-order two-dimensional (2D) chi-square ($\chi_2^2$) distribution as a null hypothesis. In particular, we tested the goodness of the model fit via a Kolmogorov-Smirnov (KS) test \citep{press2007}. When the KS test does not satisfy the null hypothesis, we adjust  the slope, $m,$ and the intercept, $b,$ slightly, until the KS test was fulfilled. So, the obtained model line indicates for which values of the regression parameters it manifests a well-defined power-law distribution. However, we do this on cost of sacrificing slightly the precision of the regression procedure. 


Afterwards, we estimated the 95\% probability limit $\gamma_e$ by assuming that the noise is 2D $\chi^2$ distributed, following Eq.~(16) from \citet{Vaughan2005}.

To cross-validate the results, we applied the above described procedures to four different methods. The first one is method M1, where we analyse the original data without applying detrendization or apodization procedures (see Fig.~\ref{method}, panel M1). The second one is method M2, where the data are pre-processed with detrendization, namely, a subtraction of the appropriate polynomial trend from the original data, so that the remaining part of the data are free of artefacts of the edge effect and satisfy the power-law dependence proven by the KS test (see Fig.~\ref{method} panel M2). The data for method M3 are apodized  (see Fig.~\ref{method} panel M3) with a Gaussian function with coefficient $\alpha=3$ (for details see Paper~II). In the last method, M4, the data are apodized and detrended at the same time (see Fig.~\ref{method} panel M4). It is important to note that, for each particular dataset we used the same detrendization order both in methods M2 and M4.
These four methods are generally used in the entire frequency range of the considered AR fluctuation spectra. However, for the more detailed study of the middle- and high-frequency part of the spectrum ($\lg f > -1$), to achieve more confidence, we also applied a multi-scale 1D wavelet analysis of the original data to reveal the background trends (instead of applying the polynomial fit). We refer to this approach as the WL method. The comparison of performance of the methods M1, M2, M3, and WL in the reduced range of frequencies is shown in Fig.~\ref{method2}.

\subsection{Cross-sectional areas }
To evaluate the cross sectional areas, $\Sigma_{th}$, of the ARs, we set the thresholds, $th,$ for the magnetic field pixel values (similarly to Paper~II), assuming the pixels with a  magnetic field strength exceeding the threshold to be part of the AR. We choose the same values, namely, $th \equiv \{$ $\pm~350\;$G, $\pm~700\;$G and $\pm~900\;$G $\}$, of the thresholds for the magnetic field in each rectangular snapshot of all five ARs. These thresholds are necessary for identifying the AR body and distinguishing it from the spurious small magnetic features around. Hence, the number $N_{th}$ of the selected 'active' pixels provides an estimate of the total area of the magnetic field concentration that exceeds the applied threshold $th$. The size of each pixel on the solar surface is set by the CEA projection mapping as $S_{CEA}=1.33\times10^5\;$km$^2$ \citep{Bobra14}. Consequently, the principal components of the AR image (i.e. the main sunspots or areas of large magnetic field concentration for the positive and negative polarity) are defined as: 
\begin{equation}\label{area}
\Sigma_{th}=N_{th} S_{CEA}.
\end{equation}
The uncertainty of the area was estimated by the method given in \citet{Saha2001}.
Altogether, we considered three basic kinds of the AR cross sectional areas $\Sigma_{th}$ defined for the positive (i.e. $\Sigma_{th}^{+}$), negative (i.e. $\Sigma_{th}^{-}$), and total (i.e. $\Sigma_{th}^{tot}$) magnetic field, specified for all the  thresholds  introduced above.
We defined the ranges of AR area absolute values of the  spectrum slopes (AVSSs) $\left |m_{\Sigma_{th}^J}\right |$ corresponding to various magnetic field thresholds in Table~\ref{table1} and Table~\ref{table2} (see a detailed description of the content of these tables below, at the start of Sect.~3) defined as:
\begin{equation}\label{defarearange}
\left [ \left |m_{\Sigma_{th}^J}\right | \right ] = \left [ \min \left \{ \left |m_{\Sigma_{th}^J}\right | \right \},\max \left \{ \left |m_{\Sigma_{th}^J} \right | \right \} \right ], 
\end{equation}
where, $\min$ and $\max$ denote the minimum and maximum values for a given subset of $\Sigma_{th}^I$-s composed by certain combinations of the magnetic field polarity ($J\subset \{ +, -, tot\}$) and applied threshold ($th\subset \{ 350, 700, 900\}$) values. 
Values within squared brackets indicate the range of AVSSs, whereas the corresponding subset of considered cases are within curly brackets.

\subsection{Mean and total radial magnetic fluxes}
We estimated the total (and mean) radial magnetic flux of the considered ARs in Mx.
The radial magnetic flux data give the total signed (TSMF) and unsigned magnetic fluxes (TUMF) as well as the mean signed (MSMF) and unsigned (MUMF) magnetic fluxes. Also, we evaluated the total (TPMF) and mean (MPMF) positive magnetic fluxes as well as the total (TNMF) and mean (MNMF) negative magnetic fluxes. The combined expression of the various versions of the radial magnetic flux reads as follows: 
$$\Phi_{th}^{I} = \sum_{\hbox{\tiny active pixels}} B_{I}(x,y)S(x,y) = $$
\begin{equation}\label{flux}
= S_{CEA} \sum \left \{B_{I}(x,y) \;|\; B_{I}(x,y)>th \right \},
\end{equation}
where the index $th$ again indicates the applied magnetic field threshold and the index $I$ indicates the type of the considered magnetic field, $B_{I}$, for which the flux is defined as follows:
\begin{align*}
B_{I} \subset & \{ B_r ~\mbox{(i.e. TSMF)},\quad\;\;\qquad |B_r| ~\mbox{(i.e. TUMF)},\\
&B_r/N_{th} ~\mbox{(i.e. MSMF)},\qquad |B_r|/N_{th} ~\mbox{(i.e. MUMF)},\\
&B_{r+} ~\mbox{(i.e. TPMF)},\quad\;\;\qquad B_{r+}/N_{th} ~\mbox{(i.e. MPMF)},\\
&B_{r-} ~\mbox{(i.e. TNMF)},\quad\;\qquad B_{r-}/N_{th} ~\mbox{(i.e. MNMF)} \},
\end{align*}
with the signs $+$ and $-$ in the index denoting the positive and negative polarity of the magnetic field, respectively. The values $x$ and $y$ in Eq.~(\ref{flux}) are the rectangular coordinates of the active pixels. We employed the definition via Eq.~(\ref{flux}) for all the considered thresholds $th$. Furthermore, we defined the structuring of the AVSSs of the radial magnetic fluxes, $\left |m_{\Phi_{th}^{I}}\right |$, and their ranges, $\left [ \left |m_{\Phi_{th}^{I}}\right | \right ]$, in a similar manner as done for the AR areas by Eq.~(\ref{defarearange}):
\begin{equation}\label{deffluxrange}
\left [ \left |m_{\Phi_{th}^{I}}\right | \right ] = \left [ \min \left \{ \left |m_{\Phi_{th}^I}\right | \right \},\max \left \{ \left |m_{\Phi_{th}^I}\right | \right \} \right ], 
\end{equation}
where, $\min$ and $\max$ denote the minimum and maximum values for a given subset of $\Phi_{th}^I$-s composed of certain combinations of the magnetic field type ($I\subset \{ \mbox{TSMF},\mbox{TUMF},\mbox{MSMF},\mbox{MUMF},\mbox{TPMF},\mbox{MPMF},\mbox{TNMF},\mbox{MNMF}\}$) and applied threshold ($th\subset \{ 350, 700, 900\}$) values.


\begin{figure*}[ht!]
\center{\includegraphics[width=1\textwidth]{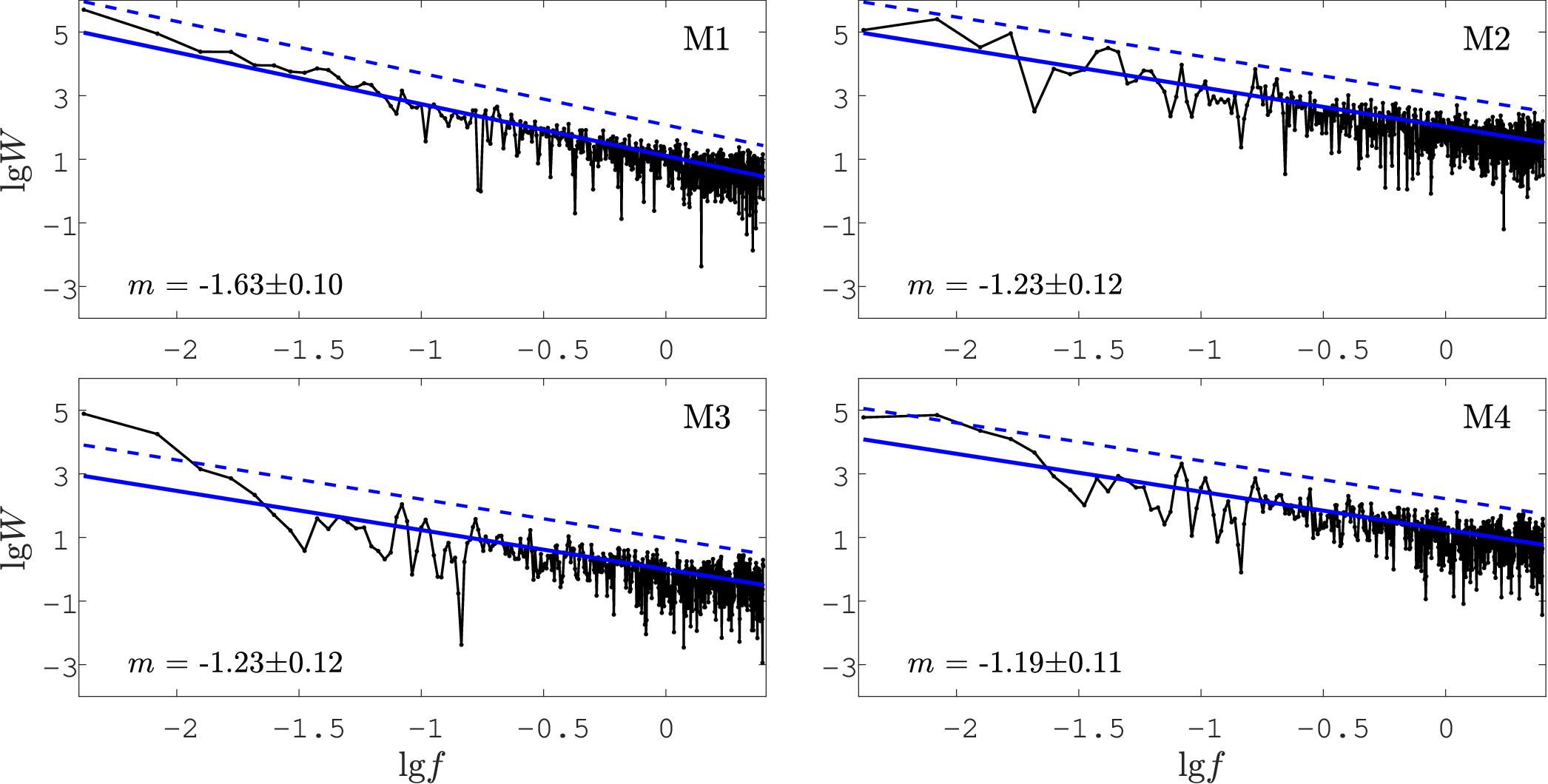}}
\caption{Comparative performance of the methods M1, M2, M3, and M4, providing the fluctuation spectra of area $\Sigma_{350}$ for AR 12524 in the entire frequency range. The particular case of threshold of $th=350\;$G is shown as an example. Here, the blue solid line indicates the model line with slope $m$ that is computed by least square fitting (Eq.~(\ref{modelline})) to the obtained power spectrum and the 95\% confidence border is represented by the blue dashed line.}
\label{method}
\end{figure*}
\begin{figure*}[ht!]
\center{\includegraphics[width=1\textwidth]{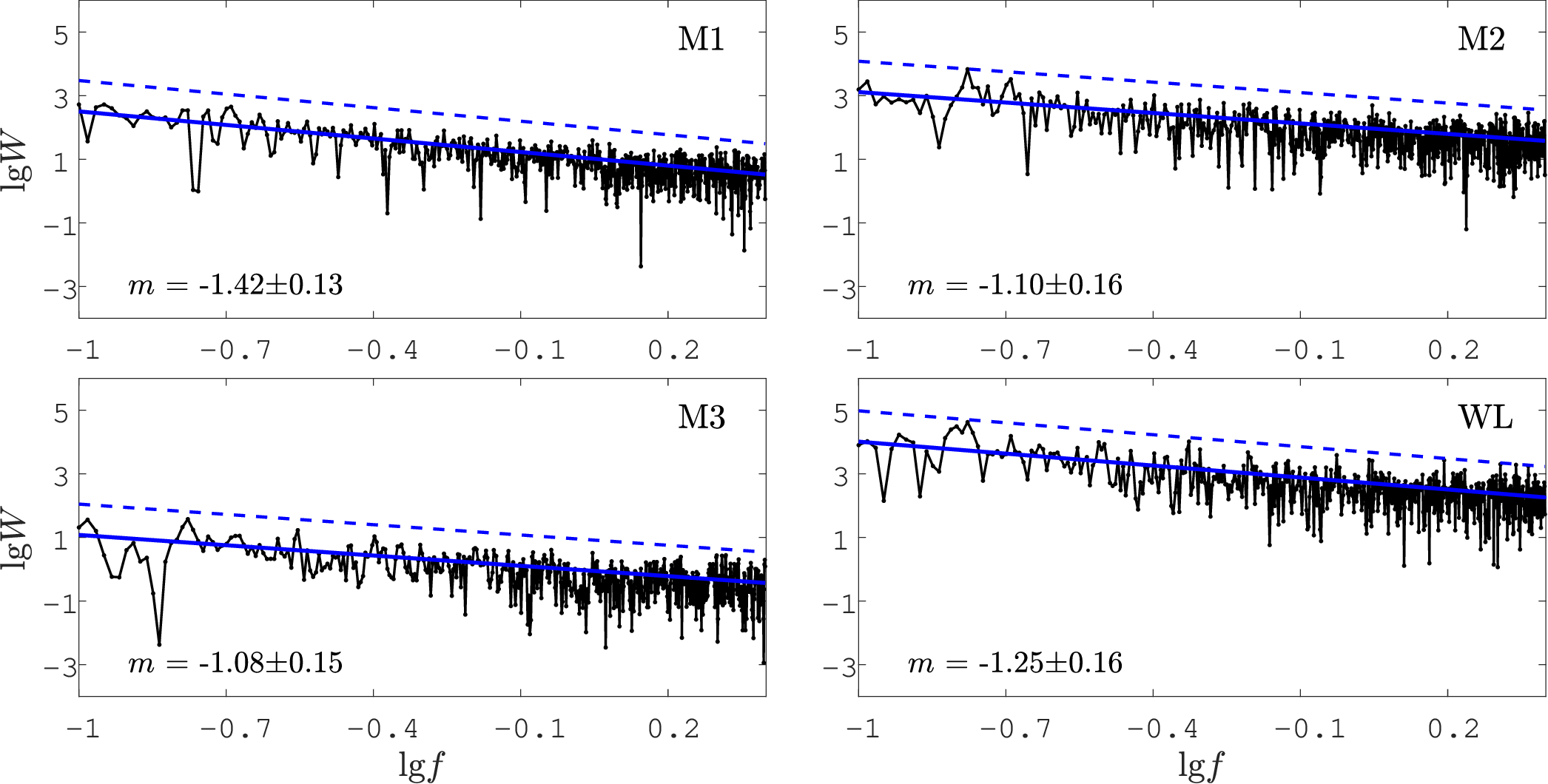}}
\caption{Spectra corresponding to case shown in Fig.~\ref{method}, obtained via approach 2.}
\label{method2}
\end{figure*}

\section{Results}
The comparison of the method M1 with the other used methods reveals that the spectra obtained with this method are drastically influenced by the dataset edge effects, which presumably create very low-frequency artefacts at the left edges of spectra constructed via approach 1. In other methods, this problem is mainly solved. Further on, we denote as approach 1 the analysis based on the methods M2, M3, and M4, which include the dedicated apodization and/or detrendization procedures. This approach was applied
across the entire frequency range (the ranges of spectral slopes obtained with
approach 1 are given in Table~\ref{table1}, with example spectra shown in
Fig.~\ref{method}). Approach 2, applied in the middle-to-high frequency range and, in addition to the M1, M2m, and M3 methods,  also includes the method WL (with the ranges of spectral slopes obtained with approach 2 are given in Table~\ref{table2} and example spectra shown in Fig.~\ref{method2}). The latter method employs the multi-scale wavelet analysis for revealing and exclusion of the background trends. Using either of these approaches enables identification and extraction of the mentioned low-frequency (presumably of artificial power) vibrations and also constructing more precise Fourier or global wavelet spectra of the remaining noise. This is necessary to obtain a realistic picture describing the nature of the fluctuation noise. In the consideration of outcomes with respect to approaches 1 and 2 below, for the sake of comparison, we will also refer to the results of M1 known to be artefact contaminated. However, the scientific conclusions will be drawn on the basis of  approaches 1 and 2 (i.e. M2, M3, M4, or WL) in the entire or middle-to-high frequency ranges, respectively.

In the subsections of Appendix~A, we describe in detail the performed analysis of the fluctuation noise spectra for each of the considered ARs. The summary of the revealed absolute values of the AVSSs for the fluctuations of all the above introduced parameters of ARs (i.e. area, TSMF, TUMF, MSMF, MUMF, TPMF, MPMF, TNMF, and MNMF). We elaborate on the different ranges of AVSSs for the constructed subsets of AR areas $\left [ \left |m_{\Sigma_{th}^J}\right | \right ]$ (defined by Eq.~(\ref{defarearange})) and magnetic field fluxes $\left [ \left |m_{\Phi_{th}^{I}}\right | \right ]$ (defined by Eq.~(\ref{deffluxrange})). In Table~\ref{table1} (approach 1) and in Table~\ref{table2} (approach 2), we also demonstrate some of the relevant subset ranges of the AVSSs, using the same definitions. In the lower-right fields of these tables, the values are absent as corresponding ARs are unipolar and, therefore, corresponding ranges of the values for these ARs, shown in rows 7, 8, and 9 are the same as in rows 4, 5, and 6, accordingly.

\begin{table*}[!ht]
\caption{AVSS ranges for some important subgroups of cases, revealed using approach 1. The colours of the AVSSs correspond to the applied magnetic field thresholds: $th=350\;$G (black); $th= 700\;$G (blue); $th=900\;$G (red).}
 \centering
\scriptsize
     \begin{threeparttable}
\begin{NiceTabular}{ccccccc}
 \hline \hline
 N &  & AR 12381 & AR 12378 & AR 12524 & AR 12435 & AR 12437 \\ \hline
$1^{\tnote{*}}$ & $\left [ \left |m _{\Sigma_{350}^{J}}\right | \right ]$ & [$1.1\pm0.11$, $1.4\pm0.11$] & [$1.1\pm0.12$, $1.3\pm0.12$] & [$1.1\pm0.11$, $1.2\pm0.12$] & [$0.8\pm0.13$, $1.1\pm0.13$] & [$0.9\pm0.12$, $1.0\pm0.12$] \\ 
\\
\RowStyle{\color{blue}}
$2^{\tnote{*}}$ & $\left [ \left |m_{\Sigma_{700}^{J}}\right | \right ]$ & [$1.3\pm0.11$, $1.6\pm0.11$] & [$1.1\pm0.12$, $1.3\pm0.13$] & [$1.2\pm0.12$, $1.5\pm0.12$] & [$1.0\pm0.12$, $1.1\pm0.12$] & [$1.2\pm0.11$, $1.3\pm0.12$] \\ 
\\
\RowStyle{\color{red}}
$3^{\tnote{*}}$ & $\left [ \left |m _{\Sigma_{900}^{J}}\right | \right ]$ & [$1.5\pm0.11$, $1.8\pm0.12$] & [$1.3\pm0.11$, $1.5\pm0.12$] & [$1.3\pm0.11$, $1.6\pm0.11$] & [$1.2\pm0.12$, $1.2\pm0.12$] & [$1.1\pm0.12$, $1.5\pm0.11$] \\  \hline
$4^{\tnote{**}}$ & $\left [ \left |m_{\Phi_{350}^{I}}\right | \right ]$ & [$1.2\pm0.11$, $1.6\pm0.12$] & [$1.2\pm0.12$, $1.3\pm0.12$] & [$1.1\pm0.11$, $1.2\pm0.11$] & [$0.9\pm0.13$, $1.2\pm0.13$] & [$0.9\pm0.13$, $1.1\pm0.12$] \\ 
\\
\RowStyle{\color{blue}}
$5^{\tnote{**}}$ & $\left [ \left |m_{\Phi_{700}^{I}}\right | \right ]$ & [$1.4\pm0.11$, $2.0\pm0.12$] & [$1.1\pm0.11$, $1.5\pm0.12$] & [$1.2\pm0.12$, $1.6\pm0.11$] & [$1.0\pm0.12$, $1.2\pm0.12$] & [$1.2\pm0.13$, $1.4\pm0.12$] \\ 
\\
\RowStyle{\color{red}}
$6^{\tnote{**}}$ & $\left [ \left |m_{\Phi_{900}^{I}}\right | \right ]$ & [$1.6\pm0.11$, $2.0\pm0.12$] & [$1.3\pm0.11$, $1.5\pm0.12$] & [$1.4\pm0.12$, $1.8\pm0.12$] & [$1.2\pm0.12$, $1.3\pm0.11$] & [$1.1\pm0.12$, $1.5\pm0.11$] \\ \hline
$7^{\tnote{***}}$ & $\left [ \left |m_{\Phi_{350}^{I}}\right | \right ]$ & [$1.1\pm0.12$, $1.4\pm0.13$] & [$1.1\pm0.12$, $1.3\pm0.13$] & [$1.1\pm0.11$, $1.3\pm0.12$] & -- & -- \\ 
\\
\RowStyle{\color{blue}}
$8^{\tnote{***}}$ & $\left [ \left |m_{\Phi_{700}^{I}}\right | \right ]$ & [$1.3\pm0.11$, $1.8\pm0.11$] & [$1.1\pm0.12$, $1.3\pm0.12$] & [$1.2\pm0.12$, $1.5\pm0.12$] & -- & -- \\ 
\\
\RowStyle{\color{red}}
$9^{\tnote{***}}$  & $\left [ \left |m_{\Phi_{900}^{I}}\right | \right ]$ & [$1.5\pm0.11$, $1.9\pm0.11$] & [$1.3\pm0.12$, $1.6\pm0.11$] & [$1.3\pm0.12$, $1.8\pm0.12$] & -- & -- \\ \hline
\end{NiceTabular} \label{table1}
    \begin{tablenotes}
        \item[*] $J\subseteq \{ +, -, tot\}$.
        \item[**] $I\subseteq \{ \mbox{TPMF}, \mbox{MPMF}, \mbox{TNMF}, \mbox{MNMF} \}$.
        \item[***] $I\subseteq \{ \mbox{TSMF},\mbox{TUMF}, \mbox{MSMF}, \mbox{MUMF} \}$.
    \end{tablenotes}
    \end{threeparttable}
\end{table*}

\begin{table*}[!ht]
\caption{Same as in Table~\ref{table1}, obtained by approach 2. }
 \centering
\scriptsize
     \begin{threeparttable}
\begin{NiceTabular}{ccccccc}
 \hline \hline
N & & AR 12381 & AR 12378 & AR 12524 & AR 12435 & AR 12437 \\ \hline
$1^{\tnote{*}}$ & $\left [ \left |m _{\Sigma_{350}^{J}}\right | \right ]$ & [$0.9\pm0.16$, $1.3\pm0.15$] & [$0.9\pm0.16$, $1.2\pm0.17$] & [$1.0\pm0.15$, $1.3\pm0.16$] & [$0.7\pm0.17$, $1.1\pm0.17$] & [$0.8\pm0.17$, $0.9\pm0.17$] \\ 
\\
\RowStyle{\color{blue}}
$2^{\tnote{*}}$ & $\left [ \left |m _{\Sigma_{700}^{J}}\right | \right ]$ & [$1.1\pm0.15$, $1.4\pm0.15$] & [$0.9\pm0.15$, $1.3\pm0.16$] & [$1.1\pm0.15$, $1.4\pm0.15$] & [$1.0\pm0.16$, $1.0\pm0.16$] & [$1.1\pm0.15$, $1.4\pm0.17$] \\ 
\\
\RowStyle{\color{red}}
$3^{\tnote{*}}$ & $\left [ \left |m _{\Sigma_{900}^{J}}\right | \right ]$ & [$1.4\pm0.15$, $1.7\pm0.17$] & [$1.1\pm0.16$, $1.5\pm0.14$] & [$1.2\pm0.16$, $1.5\pm0.16$] & [$1.0\pm0.16$, $1.2\pm0.17$] & [$0.9\pm0.16$, $1.4\pm0.18$] \\ \hline
$4^{\tnote{**}}$ &   $\left [ \left |m_{\Phi_{350}^{I}}\right | \right ]$ & [$1.0\pm0.14$, $1.6\pm0.16$] & [$1.0\pm0.16$, $1.2\pm0.16$] & [$1.0\pm0.15$, $1.3\pm0.15$] & [$0.7\pm0.16$, $1.2\pm0.16$] & [$0.8\pm0.16$, $1.0\pm0.17$] \\ 
\\
\RowStyle{\color{blue}}
$5^{\tnote{**}}$ & $\left [ \left |m_{\Phi_{700}^{I}}\right | \right ]$ & [$1.2\pm0.15$, $1.9\pm0.19$] & [$1.0\pm0.15$, $1.4\pm0.15$] & [$1.1\pm0.15$, $1.5\pm0.14$] & [$1.0\pm0.17$, $1.1\pm0.16$] & [$1.1\pm0.15$, $1.4\pm0.16$] \\ 
\\
\RowStyle{\color{red}}
$6^{\tnote{**}}$ & $\left [ \left |m_{\Phi_{900}^{I}}\right | \right ]$ & [$1.4\pm0.15$, $2.0\pm0.16$] & [$1.1\pm0.16$, $1.5\pm0.14$] & [$1.2\pm0.17$, $1.6\pm0.16$] & [$1.1\pm0.16$, $1.2\pm0.18$] & [$0.9\pm0.16$, $1.4\pm0.17$] \\ \hline
$7^{\tnote{***}}$ & $\left [ \left |m_{\Phi_{350}^{I}}\right | \right ]$ & [$0.8\pm0.16$, $1.4\pm0.16$] & [$0.9\pm0.17$, $1.2\pm0.16$] & [$0.9\pm0.16$, $1.3\pm0.16$] & -- & -- \\ 
\\
\RowStyle{\color{blue}}
$8^{\tnote{***}}$ & $\left [ \left |m_{\Phi_{700}^{I}}\right | \right ]$ & [$1.2\pm0.15$, $1.8\pm0.15$] & [$1.0\pm0.15$, $1.3\pm0.16$] & [$1.2\pm0.16$, $1.4\pm0.17$] & -- & -- \\ 
\\
\RowStyle{\color{red}}
$9^{\tnote{***}}$ & $\left [ \left |m_{\Phi_{900}^{I}}\right | \right ]$ & [$1.4\pm0.15$, $1.9\pm0.15$] & [$1.2\pm0.16$, $1.6\pm0.15$] & [$1.2\pm0.15$, $1.7\pm0.16$] & -- & -- \\
 \hline
\end{NiceTabular} \label{table2}
    \begin{tablenotes}
        \item[*] $J\subseteq \{ +, -, tot\}$.
        \item[**] $I\subseteq \{ \mbox{TPMF}, \mbox{MPMF}, \mbox{TNMF}, \mbox{MNMF} \}$.
        \item[***] $I\subseteq \{ \mbox{TSMF},\mbox{TUMF}, \mbox{MSMF}, \mbox{MUMF} \}$.
    \end{tablenotes}
    \end{threeparttable}
\end{table*}

\begin{figure*}[!ht]
\centering
\includegraphics[width=0.8\textwidth]{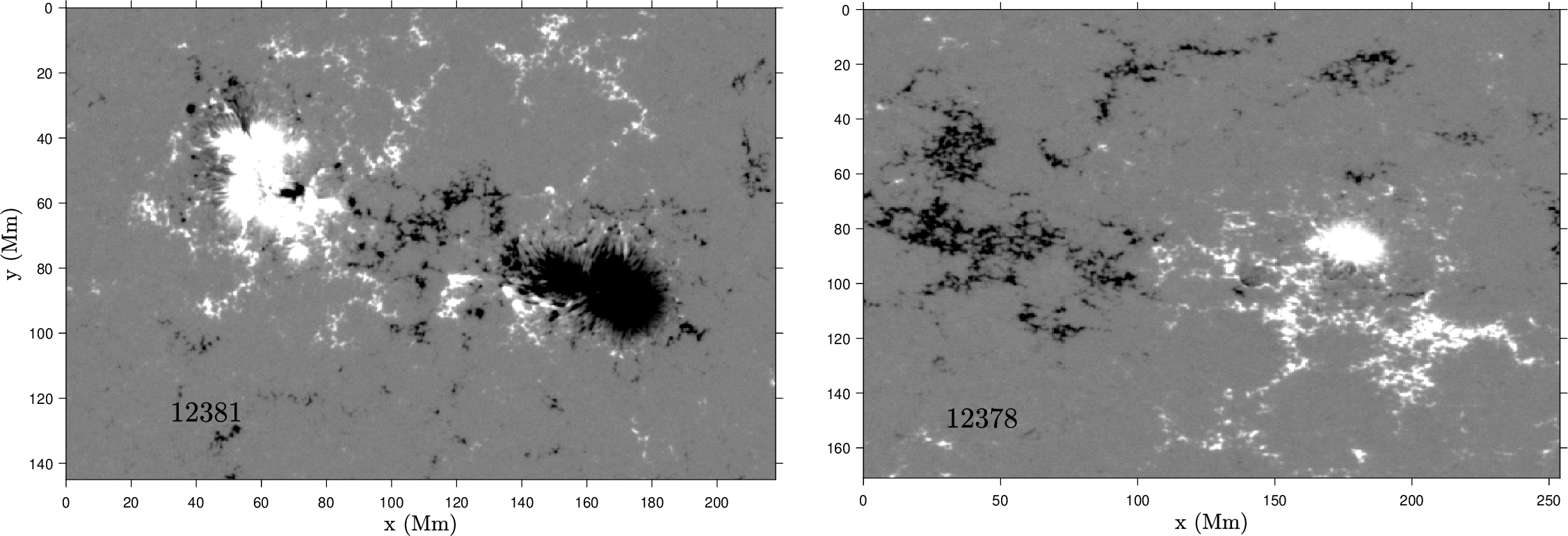}
\caption{Example illustration of Helioseismic and Magnetic Imager (HMI)  magnetogram snapshots of the compact (AR 12381) and dispersed (AR 12378) active regions. }
\label{ARexample}
\end{figure*}

The results of performed analysis of the fluctuation noise spectra presented in this work can be summarised as follows:

\begin{itemize}
    \item {For the AR areas fluctuations: 
\\[0.2cm]    
(A) The specific feature found for the area fluctuation spectra of the compact AR 12381 is that the AVSSs of the negative magnetic field area are always larger than those in the case of the positive magnetic field. Such an interrelation between these features is not seen in the spectra of the dispersed, mixed, and unipolar ARs studied (see subsection~\ref{AR_12381} point (i)). 
\\[0.2cm]
(B) In the compact AR 12381, the AVSSs ranges are relatively wider as compared to other considered cases, and they strongly depend on the applied magnetic field threshold (see subsection~\ref{AR_12381} points (i) and (ii)). However, this dependence on the magnetic field threshold is weaker for all other ARs. 

The AVSSs for the dispersed type AR 12378 coincide with those for the compact AR 12381 in the case of the relatively small magnetic fields threshold ($350\;$G); while for the thresholds ($700\;$G and $900\;$G), they become significantly smaller (see subsection~\ref{AR_12378} points (i) and (ii)). 

The mixed AR 12524 comprises the properties of both the compact and dispersed ARs with the maximal AVSSs so that their ranges for the applied large magnetic field thresholds somewhat exceed the corresponding values measured for the dispersed AR 12378. At the same time, they still remain significantly smaller than those for the compact AR 12381 (see subsection~\ref{AR_12524} points i and ii). 

The character of area fluctuations in the case of two unipolar ARs, namely, AR 12435 and AR 12437, is similar to that of the dispersed AR 12378. Moreover, the narrowing of the AVSSs' ranges, depending on the applied magnetic field threshold, is more pronounced for the large thresholds (see Section~\ref{AR_12435} points i and ii).
\\[0.2cm]
(C) The ranges of AVSSs for the AR area fluctuation spectra, obtained with the $350\;$G threshold, are close to those typical for the $1/f-$ type spectrum. This is the case for all ARs under study (see subsections~\ref{AR_12381}, \ref{AR_12378}, \ref{AR_12524}, and \ref{AR_12435} points (ii)). 

For the higher magnetic field thresholds applied, the AR area fluctuation noise spectra for the dispersed bipolar (AR 12378) and two dispersed unipolar (AR 12435 and AR 12437) ARs remain predominantly close to the $1/f-$ type one. At the same time, for the compact AR 12381 and the mixed-type AR 12524, the AR area fluctuation noise spectra manifest the presence of a pink noise for the $700\;$G threshold, whereas for the $900\;$G threshold, they become inclined towards the random walk red noise type.

Altogether, the systematic interrelation between the AVSSs of the AR area fluctuations for the positive, negative, and total magnetic fields does not depend on the considered total area of the corresponding AR parts, nor on which parts are leading or trailing. The observed interrelations between the AVSSs in the case of compact AR 12381 might be the result of its specific internal magnetic field structure, but this assumption needs further studies.    
} 
\\[0.5cm] 

\item {For the AR magnetic flux fluctuations: 
\\[0.2cm]
(A) Similarly to the ARs' area fluctuations, the AVSSs for the negative magnetic flux in the compact AR always exceed those for the positive magnetic flux  (see Section~\ref{AR_12381}, point iii). At the same time, in all the other ARs such a systematic relation between the corresponding AVSSs is not observed (see Sections~\ref{AR_12378}, \ref{AR_12524}, and \ref{AR_12435}, point iii). 
\\[0.2cm]
(B) Similarly to the case of the area fluctuation spectra, the ranges of AVSS for the magnetic fluxes of the compact AR 12381 are relatively wide and strongly dependent on the applied magnetic field threshold. However, in this case, the AVSSs for the magnetic fluxes of different polarities are larger than those for the corresponding area fluctuation spectra (see subsection~\ref{AR_12381} point (v)). 

For the fully or partially dispersed ARs, the AVSS ranges are rather narrow and the observed dependence on the magnetic field threshold is weakly pronounced as compared to the case of the compact AR 12381 (see Sections~\ref{AR_12378}, \ref{AR_12524}, and \ref{AR_12435} points (v)). In the latter cases, the estimated AVSSs for the magnetic fluxes of the different polarities are evaluated within the ranges with widths comparable to those in the cases of the area fluctuation spectra. However, the difference between them consists of the different maximum and minimum values of the AVSSs. Although the mixed AR 12524 spectra partially comprise the features of the compact and dispersed ARs. 
\\[0.2cm]
(C) For the compact AR 12381 (see the snapshots of this AR in left panel of Fig.~\ref{ARexample}), the pink spectrum is observed for the threshold of $350\;$G, with the AVSSs values still close to the $1/f-$ type spectrum. Nevertheless, they are not of a strictly $1/f-$ type, as in the case of the area fluctuations. With regard to the $700\;$G and $900\;$G thresholds, the corresponding AVSSs are also typical for the pink spectrum, with more inclination towards the red one. This might indicate the presence of a random walk component in the magnetic flux fluctuations (see Section~\ref{AR_12381}, point v). 

In the dispersed AR 12378 (see the snapshots of this AR in right panel of Fig.~\ref{ARexample}) and the unipolar AR 12435 and AR 12437, we have strictly defined $1/f-$type spectra for all magnetic field thresholds (see Sections~\ref{AR_12378} and \ref{AR_12435}, points v). The situation is similar to the case of the mixed AR 12524. However, the typical values of AVSSs are slightly larger, which is probably due to the presence of a compact magnetic kernel in this AR (see Section~\ref{AR_12524}, point v).
\\[0.2cm]
(D) 
The general result for the compact AR 12381 is that the AVSS ranges of the signed and unsigned magnetic flux fluctuations are wider than those of the spectra of different polarities, but still strongly dependent on the applied magnetic field threshold, similarly to other cases of spectra constructed for this AR. Again, the estimated AVSSs for the signed and unsigned magnetic fluxes are (in most cases) larger than those for the corresponding area fluctuation spectra (see Section~\ref{AR_12381}, point vi). In addition, in the compact AR, the AVSSs for the unsigned magnetic flux are always larger than those of the signed (uncompensated) magnetic flux (see Section~\ref{AR_12381}, point iv). In all other ARs, we do not observe such a feature (see subsections~\ref{AR_12378}, \ref{AR_12524}, and \ref{AR_12435} points (iv)). 

In the dispersed AR 12378 and the mixed AR 12524, it is noticeable that the AVSSs obtained for the 700~G and 900~G thresholds for the total unsigned magnetic flux (triangles) appear to be larger than the AVSSs in most of the other spectra. This might indicate that a random walk process of the magnetic flux emergence or cancellation is more strongly present in the ARs of these types.
}
\end{itemize}

\section{Conclusions}

The case study performed in this work for the ARs' areas and magnetic flux fluctuation spectra leads us to the conclusion that they are different for compact and dispersed type ARs, which can be an indication of the dependence of these spectra on the distinct spatially random internal structure and topology of the magnetic field. However, the fact that the AVSSs for the magnetic fluxes are somewhat larger than the corresponding values for the ARs' area fluctuations may manifest the presence of the processes (presumably of a statistically non-equilibrium nature) of the magnetic flux emergence or cancellation. The evident differences between the area and magnetic flux spectra can indicate that the background noise of these physical quantities may be driven by the spatiotemporal processes of different physical natures. Further analytical modelling
and data analysis procedures are needed to improve our understanding of the relative importance between spatially random magnetic field structure and temporarily stochastic magnetic flux variability in the formation of the observed fluctuation spectra.

\begin{acknowledgements}
We thank the referee for constructive remarks that led to a significant improvement of the content.
The work of G.D. was partially supported under the DAAD-SRNSFG Scholarships for Young Scientists, 2023 provided by the Deutscher Akademischer Austauschdienst (DAAD) and Shota Rustaveli National Science Foundation of Georgia (SRNSFG) - project Ref. ID. 91881041. We also acknowledge partial financial support for B.S.\ from the Ruhr-Universit\"at Bochum (RUB) with in the VIP programme and from the Deutscher Akademischer Austauschdienst (DAAD) within EU fellowships for Georgian researchers, 2023 (57655523) - project Ref.\ ID.\ 91862684.  SP acknowledges support from the projects C14/19/089  (C1 project Internal Funds KU Leuven), G.0B58.23N and G.0025.23N (WEAVE)   (FWO-Vlaanderen), 4000134474 (SIDC Data Exploitation, ESA Prodex-12), and Belspo project B2/191/P1/SWiM.
\end{acknowledgements}

\bibliography{mybib}

\begin{thebibliography}{30}
\expandafter\ifx\csname natexlab\endcsname\relax\def\natexlab#1{#1}\fi

\bibitem[{{Auch{\`e}re} {et~al.}(2014){Auch{\`e}re}, {Bocchialini}, {Solomon},
  \& {Tison}}]{Auchere2014}
{Auch{\`e}re}, F., {Bocchialini}, K., {Solomon}, J., \& {Tison}, E. 2014, \aap,
  563, A8

\bibitem[{{Auch{\`e}re} {et~al.}(2016){Auch{\`e}re}, {Froment}, {Bocchialini},
  {Buchlin}, \& {Solomon}}]{Auchere2016}
{Auch{\`e}re}, F., {Froment}, C., {Bocchialini}, K., {Buchlin}, E., \&
  {Solomon}, J. 2016, \apj, 825, 110

\bibitem[{{Bobra} {et~al.}(2014){Bobra}, {Sun}, {Hoeksema}, {Turmon}, {Liu},
  {Hayashi}, {Barnes}, \& {Leka}}]{Bobra14}
{Bobra}, M.~G., {Sun}, X., {Hoeksema}, J.~T., {et~al.} 2014, \solphys, 289,
  3549

\bibitem[{{Delaboudini{\`e}re} {et~al.}(1995){Delaboudini{\`e}re}, {Artzner},
  {Brunaud}, {Gabriel}, {Hochedez}, {Millier}, {Song}, {Au}, {Dere}, {Howard},
  {Kreplin}, {Michels}, {Moses}, {Defise}, {Jamar}, {Rochus}, {Chauvineau},
  {Marioge}, {Catura}, {Lemen}, {Shing}, {Stern}, {Gurman}, {Neupert},
  {Maucherat}, {Clette}, {Cugnon}, \& {van Dessel}}]{Delaboudiniene1995}
{Delaboudini{\`e}re}, J.~P., {Artzner}, G.~E., {Brunaud}, J., {et~al.} 1995,
  \solphys, 162, 291

\bibitem[{{Domingo} {et~al.}(1995){Domingo}, {Fleck}, \&
  {Poland}}]{Domingo1995}
{Domingo}, V., {Fleck}, B., \& {Poland}, A.~I. 1995, \solphys, 162, 1

\bibitem[{{Dumbadze} {et~al.}(2017){Dumbadze}, {Shergelashvili}, {Kukhianidze},
  {Ramishvili}, {Zaqarashvili}, {Khodachenko}, {Gurgenashvili}, {Poedts}, \&
  {De Causmaecker}}]{Dumbadze17}
{Dumbadze}, G., {Shergelashvili}, B.~M., {Kukhianidze}, V., {et~al.} 2017,
  \aap, 597, A93

\bibitem[{{Dumbadze} {et~al.}(2021){Dumbadze}, {Shergelashvili}, {Poedts},
  {Zaqarashvili}, {Khodachenko}, \& {De Causmaecker}}]{Dumbadze21}
{Dumbadze}, G., {Shergelashvili}, B.~M., {Poedts}, S., {et~al.} 2021, \aap,
  653, A39

\bibitem[{{Efremov} {et~al.}(2014){Efremov}, {Parfinenko}, {Solov'ev}, \&
  {Kirichek}}]{Efremov2014}
{Efremov}, V.~I., {Parfinenko}, L.~D., {Solov'ev}, A.~A., \& {Kirichek}, E.~A.
  2014, \solphys, 289, 1983

\bibitem[{{Efremov} {et~al.}(2018){Efremov}, {Solov'ev}, {Parfinenko},
  {Riehokainen}, {Kirichek}, {Smirnova}, {Varun}, {Bakunina}, \&
  {Zhivanovich}}]{Efremov2018}
{Efremov}, V.~I., {Solov'ev}, A.~A., {Parfinenko}, L.~D., {et~al.} 2018, \apss,
  363, 61

\bibitem[{{Foullon} {et~al.}(2004){Foullon}, {Verwichte}, \&
  {Nakariakov}}]{Foullon2004}
{Foullon}, C., {Verwichte}, E., \& {Nakariakov}, V.~M. 2004, \aap, 427, L5

\bibitem[{{Foullon} {et~al.}(2009){Foullon}, {Verwichte}, \&
  {Nakariakov}}]{foullon09}
{Foullon}, C., {Verwichte}, E., \& {Nakariakov}, V.~M. 2009, \apj, 700, 1658

\bibitem[{{Froment} {et~al.}(2015){Froment}, {Auch{\`e}re}, {Bocchialini},
  {Buchlin}, {Guennou}, \& {Solomon}}]{Froment2015}
{Froment}, C., {Auch{\`e}re}, F., {Bocchialini}, K., {et~al.} 2015, \apj, 807,
  158

\bibitem[{{Gupta}(2014)}]{Gupta2014}
{Gupta}, G.~R. 2014, \aap, 568, A96

\bibitem[{{Ireland} {et~al.}(2015){Ireland}, {McAteer}, \&
  {Inglis}}]{Ireland2015}
{Ireland}, J., {McAteer}, R.~T.~J., \& {Inglis}, A.~R. 2015, \apj, 798, 1

\bibitem[{{Kolotkov} {et~al.}(2016){Kolotkov}, {Anfinogentov}, \&
  {Nakariakov}}]{Kolotkov2016}
{Kolotkov}, D.~Y., {Anfinogentov}, S.~A., \& {Nakariakov}, V.~M. 2016, \aap,
  592, A153

\bibitem[{{Kolotkov} {et~al.}(2017){Kolotkov}, {Smirnova}, {Strekalova},
  {Riehokainen}, \& {Nakariakov}}]{Kolotkov2017}
{Kolotkov}, D.~Y., {Smirnova}, V.~V., {Strekalova}, P.~V., {Riehokainen}, A.,
  \& {Nakariakov}, V.~M. 2017, \aap, 598, L2

\bibitem[{{Lemen} {et~al.}(2012){Lemen}, {Title}, {Akin}, {Boerner}, {Chou},
  {Drake}, {Duncan}, {Edwards}, {Friedlaender}, {Heyman}, {Hurlburt}, {Katz},
  {Kushner}, {Levay}, {Lindgren}, {Mathur}, {McFeaters}, {Mitchell}, {Rehse},
  {Schrijver}, {Springer}, {Stern}, {Tarbell}, {Wuelser}, {Wolfson}, {Yanari},
  {Bookbinder}, {Cheimets}, {Caldwell}, {Deluca}, {Gates}, {Golub}, {Park},
  {Podgorski}, {Bush}, {Scherrer}, {Gummin}, {Smith}, {Auker}, {Jerram},
  {Pool}, {Soufli}, {Windt}, {Beardsley}, {Clapp}, {Lang}, \&
  {Waltham}}]{Lemen2012}
{Lemen}, J.~R., {Title}, A.~M., {Akin}, D.~J., {et~al.} 2012, \solphys, 275, 17

\bibitem[{{Leonardis} {et~al.}(2012){Leonardis}, {Chapman}, \&
  {Foullon}}]{Leonardis2012}
{Leonardis}, E., {Chapman}, S.~C., \& {Foullon}, C. 2012, \apj, 745, 185

\bibitem[{{Maes} {et~al.}(2009){Maes}, {Neto{\v{c}}n{\'y}}, \&
  {Shergelashvili}}]{Maes2009}
{Maes}, C., {Neto{\v{c}}n{\'y}}, K., \& {Shergelashvili}, B.~M. 2009, \pre, 80,
  011121

\bibitem[{{Nakariakov} {et~al.}(2021){Nakariakov}, {Anfinogentov}, {Antolin},
  {Jain}, {Kolotkov}, {Kupriyanova}, {Li}, {Magyar}, {Nistic{\`o}}, {Pascoe},
  {Srivastava}, {Terradas}, {Vasheghani Farahani}, {Verth}, {Yuan}, \&
  {Zimovets}}]{Nakariakov2021}
{Nakariakov}, V.~M., {Anfinogentov}, S.~A., {Antolin}, P., {et~al.} 2021, \ssr,
  217, 73

\bibitem[{{Nakariakov} {et~al.}(2016){Nakariakov}, {Anfinogentov},
  {Nistic{\`o}}, \& {Lee}}]{Nakariakov2016}
{Nakariakov}, V.~M., {Anfinogentov}, S.~A., {Nistic{\`o}}, G., \& {Lee}, D.~H.
  2016, \aap, 591, L5

\bibitem[{{Nakariakov} {et~al.}(2022){Nakariakov}, {Kolotkov}, \&
  {Zhong}}]{Nakariakov2022}
{Nakariakov}, V.~M., {Kolotkov}, D.~Y., \& {Zhong}, S. 2022, \mnras, 516, 5227

\bibitem[{{Pesnell} {et~al.}(2012){Pesnell}, {Thompson}, \&
  {Chamberlin}}]{Pesnell2012}
{Pesnell}, W.~D., {Thompson}, B.~J., \& {Chamberlin}, P.~C. 2012, \solphys,
  275, 3

\bibitem[{{Philishvili} {et~al.}(2021){Philishvili}, {Shergelashvili},
  {Buitendag}, {Raes}, {Poedts}, \& {Khodachenko}}]{Philishvili2021}
{Philishvili}, E., {Shergelashvili}, B.~M., {Buitendag}, S., {et~al.} 2021,
  \aap, 645, A52

\bibitem[{Press {et~al.}(2007)Press, Teukolsky, Vetterling, \&
  Flannery}]{press2007}
Press, W., Teukolsky, S., Vetterling, W., \& Flannery, B. 2007, Numerical
  Recipes 3rd Edition: The Art of Scientific Computing (Cambridge University
  Press)

\bibitem[{Saha \& Udupa(2001)}]{Saha2001}
Saha, P. \& Udupa, J. 2001, IEEE Transactions on Pattern Analysis and Machine
  Intelligence, 23, 689

\bibitem[{{Scherrer} {et~al.}(2012){Scherrer}, {Schou}, {Bush}, {Kosovichev},
  {Bogart}, {Hoeksema}, {Liu}, {Duvall}, {Zhao}, {Title}, {Schrijver},
  {Tarbell}, \& {Tomczyk}}]{Scherrer12}
{Scherrer}, P.~H., {Schou}, J., {Bush}, R.~I., {et~al.} 2012, \solphys, 275,
  207

\bibitem[{{Schou} {et~al.}(2012){Schou}, {Scherrer}, {Bush}, {Wachter},
  {Couvidat}, {Rabello-Soares}, {Bogart}, {Hoeksema}, {Liu}, {Duvall}, {Akin},
  {Allard}, {Miles}, {Rairden}, {Shine}, {Tarbell}, {Title}, {Wolfson},
  {Elmore}, {Norton}, \& {Tomczyk}}]{Schou12}
{Schou}, J., {Scherrer}, P.~H., {Bush}, R.~I., {et~al.} 2012, \solphys, 275,
  229

\bibitem[{{Shergelashvili} {et~al.}(2022){Shergelashvili}, {Philishvili},
  {Buitendag}, {Poedts}, \& {Khodachenko}}]{Shergelashvili2022}
{Shergelashvili}, B.~M., {Philishvili}, E., {Buitendag}, S., {Poedts}, S., \&
  {Khodachenko}, M. 2022, \aap, 662, A30

\bibitem[{{Vaughan}(2005)}]{Vaughan2005}
{Vaughan}, S. 2005, \aap, 431, 391

\end{thebibliography}

\begin{appendix}
\section{Detailed description of the obtained spectra}
\subsection{AR 12381}\label{AR_12381} 
AR 12381 has a compactly organised magnetic field that appears in two main large sunspots, where the leading and trailing ones have the negative and positive polarities, respectively. The following features of the AR 12381 fluctuation noise spectra can be distinguished.

\begin{figure*}[ht!]
\center{\includegraphics[width=0.8\textwidth]{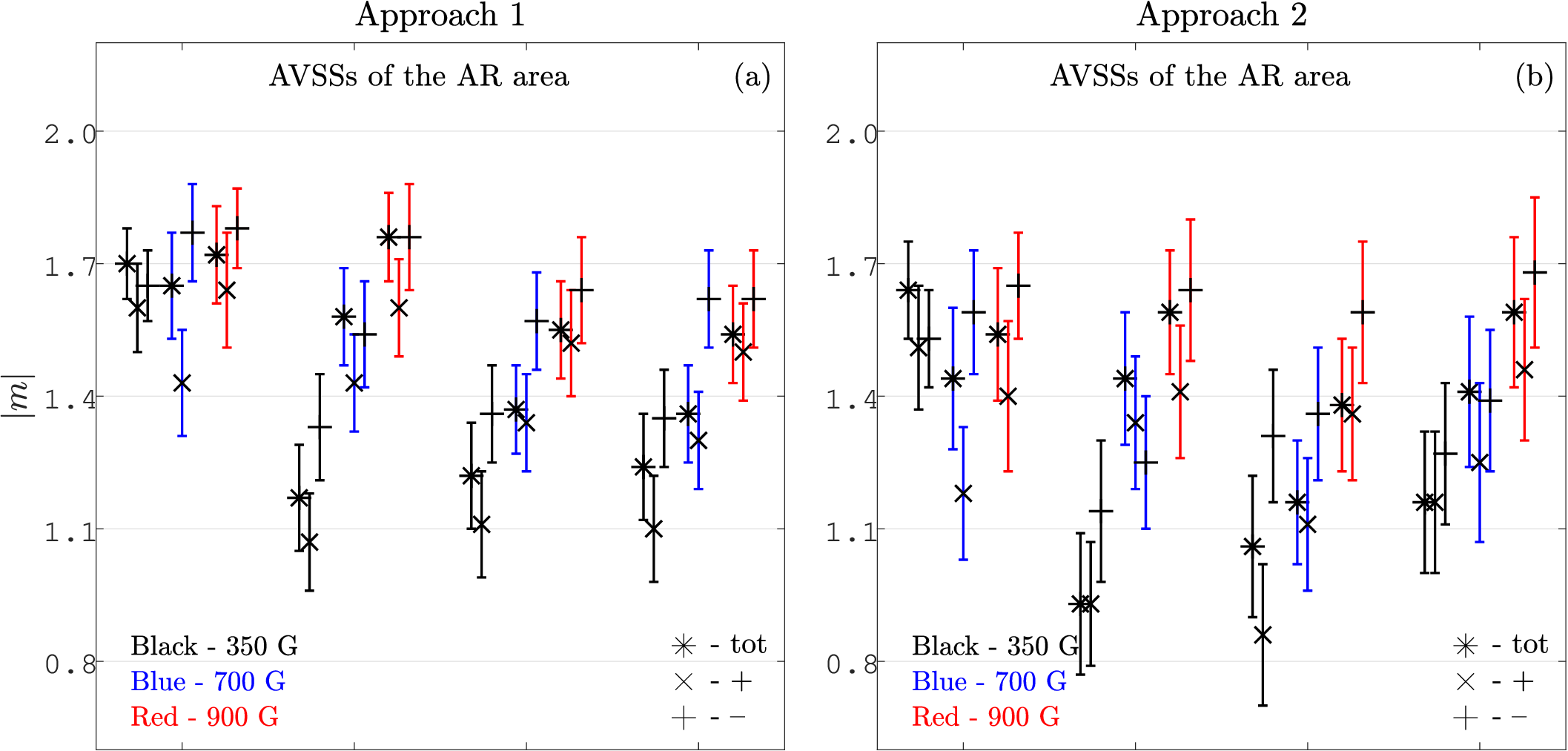}}
\center{\includegraphics[width=0.8\textwidth]{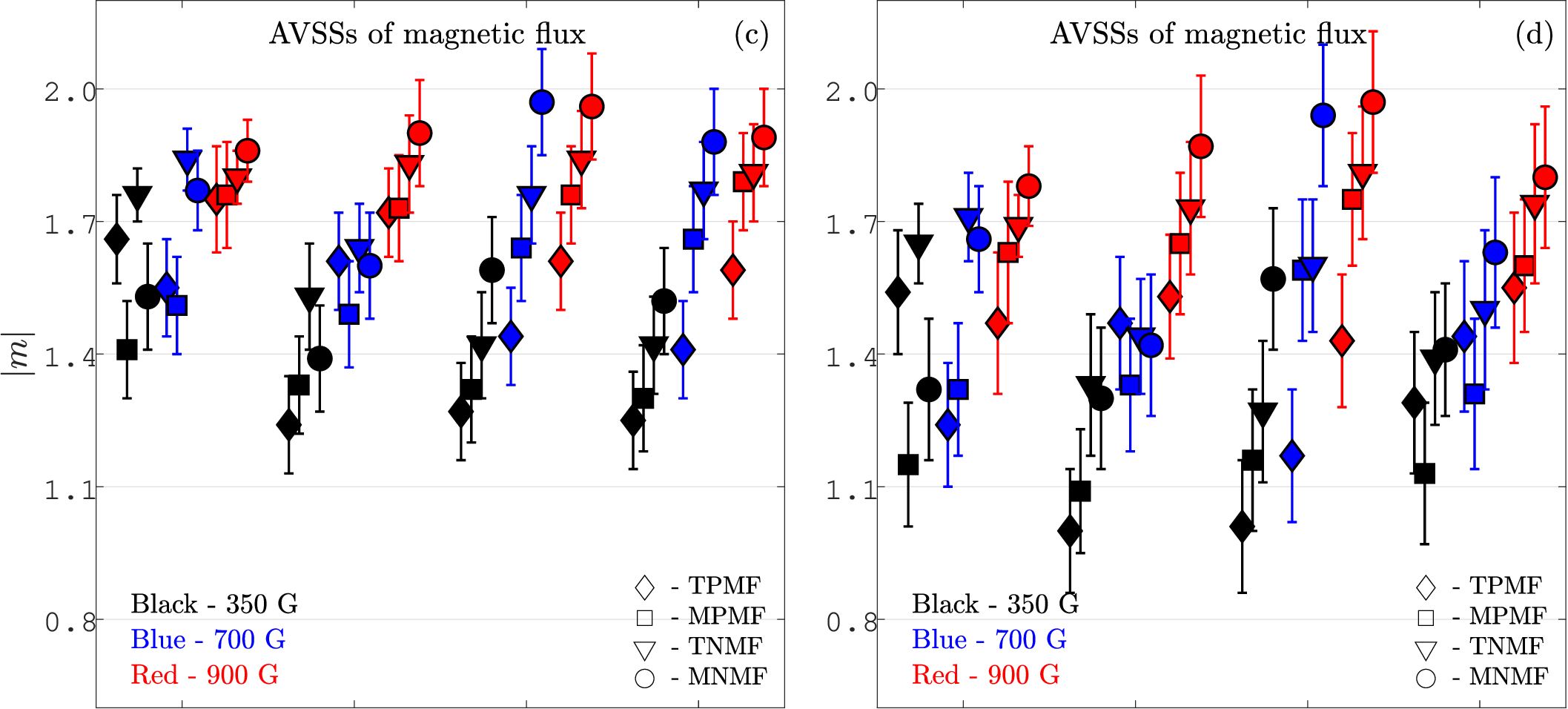}}
\center{\includegraphics[width=0.8\textwidth]{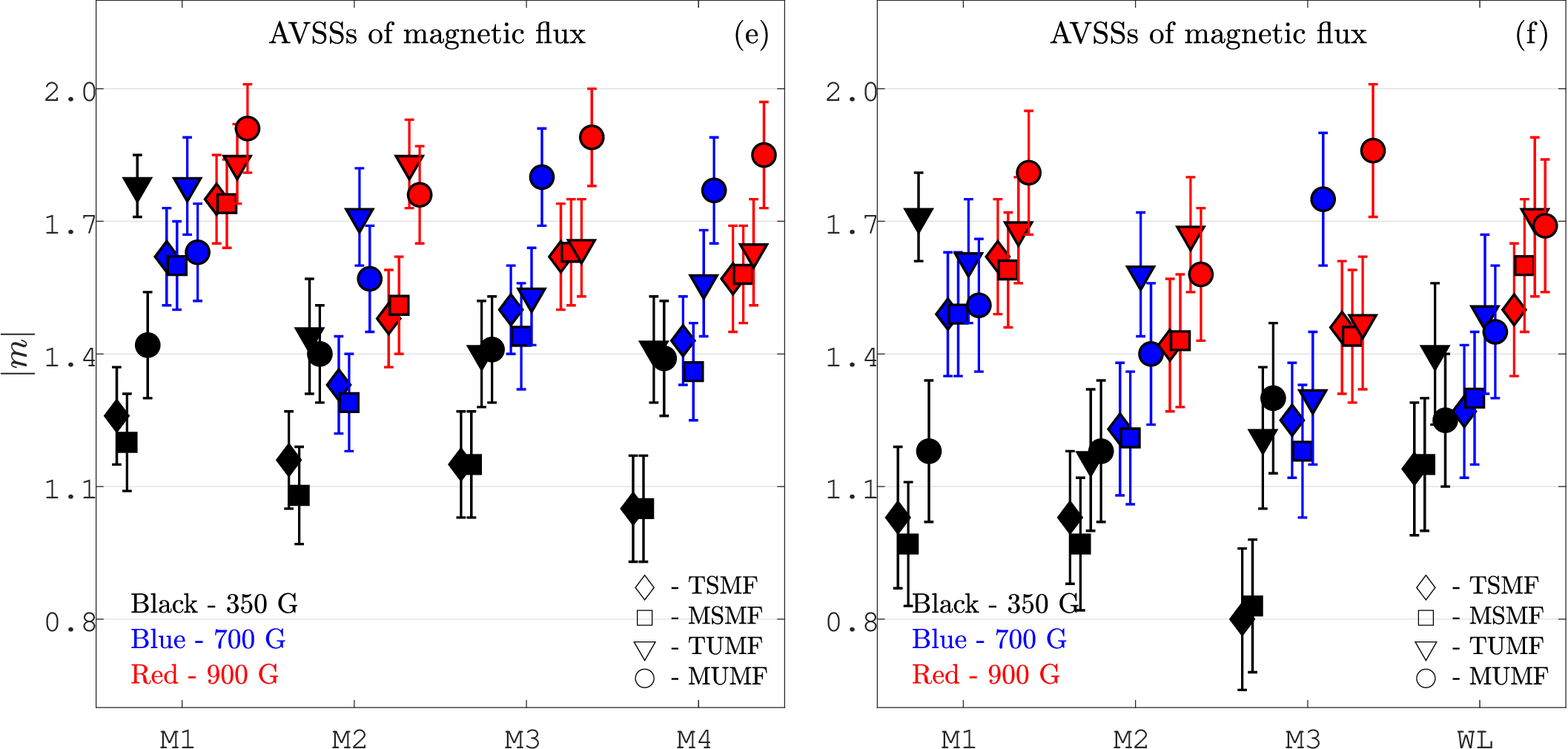}}
\caption{AVSSs obtained for the compact AR 12381 with approach 1 (left panels) and approach 2 (right panels). Panels (a) and (b) panels show the AVSSs of the AR area,  fluctuation spectra. Panels (c) and (d) present the AVSSs of the positive and negative magnetic flux fluctuation spectra. Panels (e) and (f) show the AVSSs of the signed and unsigned magnetic flux fluctuation spectra. In panels (a) and (b), the markers '$+$' and '$\times$' indicate $\left |m _{\Sigma_{th}^{+}}\right |$ and $\left |m _{\Sigma_{th}^{-}}\right |$, respectively; whereas marker '$\ast$' denotes $\left |m _{\Sigma_{th}^{tot}}\right |$ (i.e. the combination of both polarities). In panels (c) and (d), the markers '$\diamond$' and '$\square$' indicate $\left |m _{\Sigma_{th}^{TPMF, MPMF}}\right |$, while the markers '$\triangle$' and '$\circ$' are used for $\left |m _{\Sigma_{th}^{TNMF, MNMF}}\right |$. In panels (e) and (f), markers '$\diamond$' and '$\square$' show $\left |m _{\Sigma_{th}^{TSMF, MSMF}}\right |$, while markers '$\triangle$' and '$\circ$' denote $\left |m _{\Sigma_{th}^{TUMF, MUMF}}\right |$, $th\subseteq \{ 350,700,900\}$. The colours of the markers correspond to the applied magnetic field thresholds: $th=350\;$G (black); $th= 700\;$G (blue); $th=900\;$G (red). We note that for the sake of better visualisation, the markers in all plots are manually slightly shifted along the abscissa to avoid their overlapping. }
\label{results_12381}
\end{figure*}

(i) The AVSSs of the AR area fluctuations for the negative magnetic field $\left |m _{\Sigma_{th}^{-}}\right |$ (Fig.~\ref{results_12381} (a), (b), pluses) are systematically larger than those for the positive magnetic field $\left |m _{\Sigma_{th}^{+}}\right |$ (Fig.~\ref{results_12381} (a),(b), crosses). Here and further on, if the particular value of the applied magnetic field threshold $th$, magnetic field polarity and type or applied approach of analysis is not specified, then we assume that the corresponding statement holds true for all the thresholds, magnetic field cases or both approaches, accordingly.

(ii) The spectral analysis (Fig.~\ref{results_12381}, a) shows that the AVSSs for the full set of AR areas, $\left |m _{\Sigma_{th}^{J}}\right |$, $th\subseteq \{ 350,700,900\}$, $J\subseteq \{ +, -, tot\}$, fluctuation noise range within [$1.1\pm0.11$, $1.8\pm0.12$] by  approach 1, while  approach 2 (Fig.~\ref{results_12381}, b) yields the AVSSs within [$0.9\pm0.16$, $1.7\pm0.17$]. The AVSSs for the area fluctuations systematically increase with the applied magnetic field threshold $th$ in this AR (see Table~\ref{table1} and \ref{table2} rows 1,2,3). Accordingly, for $th = 350$~G, the spectrum is close to the $1/f-$ type one, with ranging within $\left [ \left |m _{\Sigma_{350}^{J}}\right | \right ]$ = [$0.9\pm0.16$, $1.4\pm0.11$], $J\subseteq \{ +, -, tot\}$ (black marks). For $th= 700$~G threshold, the well manifested pink noise spectrum with ranging within $\left [ \left |m _{\Sigma_{700}^{J}}\right | \right ]$ = [$1.1\pm0.15$, $1.6\pm0.11$], $J\subseteq \{ +, -, tot\}$ (blue marks) is seen. Application of the threshold $th = 900$~G yields the spectra with power-law inclined towards the red noise type with ranging within $\left [ \left |m _{\Sigma_{900}^{J}}\right | \right ]$ = [$1.4\pm0.15$, $1.8\pm0.12$], $J\subseteq \{ +, -, tot\}$ (red marks). 

(iii) As shown in Fig.~\ref{results_12381} (c) and (d), the AVSSs in this AR, for both, the total $\left |m_{\Phi_{th}^{TNMF}}\right |$ (triangles) and mean $\left |m_{\Phi_{th}^{MNMF}}\right |$ (circles) negative magnetic fluxes of the leading spot are in the respective order larger than those for the total $\left |m_{\Phi_{th}^{TPMF}}\right |$ (diamonds) and mean $\left |m_{\Phi_{th}^{MPMF}}\right |$ (squares) positive magnetic fluxes in the trailing spot. 

(iv) Similarly to the case of different polarities, the AVSSs of both the unsigned total $\left |m_{\Phi_{th}^{TUMF}}\right |$ (triangles) and mean $\left |m_{\Phi_{th}^{MUMF}}\right |$ (circles) magnetic flux spectra are also almost always (the exception is the case of $th=900$ G in method M3) higher when respectively compared to those of the signed total $\left |m_{\Phi_{th}^{TSMF}}\right |$ (diamonds) and mean $\left |m_{\Phi_{th}^{MSMF}}\right |$ (squares) magnetic flux (Fig.~\ref{results_12381} (e) and (f)).

(v) The spectral analysis by  approach 1 (Fig.~\ref{results_12381}, c)shows that the AVSSs of fluctuation noise for the subset of the ARs magnetic fluxes $\left |m_{\Phi_{th}^{I}}\right |$, $th\subseteq \{ 350,700,900\}$, $I\subseteq \{ \mbox{TPMF}, \mbox{MPMF}, \mbox{TNMF}, \mbox{MNMF} \}$, range within [$1.2\pm0.11$, $2.0\pm0.12$], while  approach 2 (Fig.~\ref{results_12381}, d) yields the AVSSs for the same subset within [$1.0\pm0.14$, $2.0\pm0.16$]. 
The AVSSs for the fluctuation spectra of magnetic flux of both polarities systematically increase with the applied magnetic field threshold $th$ in this AR (see Table~\ref{table1} and \ref{table2} rows 4,5,6). Accordingly, for $th= 350$ G, the spectrum is close to the $1/f-$ type one, with ranging within $\left [ \left |m_{\Phi_{350}^{I}}\right | \right ]$ = [$1.0\pm0.14$, $1.6\pm0.12$], $I\subseteq \{ \mbox{TPMF}, \mbox{MPMF}, \mbox{TNMF}, \mbox{MNMF} \}$ (black marks). For $th= 700$ G and $th= 900$ G, the well manifested pink noise spectra inclined towards the red noise are seen, with $\left [ \left |m_{\Phi_{700}^{I}}\right | \right ]$ = [$1.2\pm0.15$, $2.0\pm0.12$] (blue marks), and $\left [ \left |m_{\Phi_{900}^{I}}\right | \right ]$ = [$1.4\pm0.15$, $2.0\pm0.16$] (red marks), $I\subseteq \{ \mbox{TPMF}, \mbox{MPMF}, \mbox{TNMF}, \mbox{MNMF} \}$, respectively.

(vi) The situation is similar in the cases of the signed and unsigned magnetic flux fluctuation spectra. In particular,  approach 1 (Fig.~\ref{results_12381}, e) reveals the AVSSs for the fluctuation noise of the signed and unsigned magnetic flux $\left |m_{\Phi_{th}^{I}}\right |$, $th\subseteq \{ 350,700,900\}$, $I\subseteq \{ \mbox{TSMF}, \mbox{MSMF}, \mbox{TUMF}, \mbox{MUMF} \}$, fall in the range within [$1.1\pm0.12$, $1.9\pm0.11$]; while with  approach 2 (Fig.~\ref{results_12381}, f), we has the AVSSs range of [$0.8\pm0.16$, $1.9\pm0.15$]. Again, the AVSSs for the signed and unsigned magnetic flux fluctuations systematically increase with the applied magnetic field threshold $th$ (see Table~\ref{table1} and \ref{table2} rows 7,8,9). In particular, for $th= 350$~G, the spectrum is close to the $1/f-$ type one, with the AVSSs for unified subset of both approaches ranging within $\left [ \left |m_{\Phi_{350}^{I}}\right | \right ]$ = [$0.8\pm0.16$, $1.4\pm0.13$], $I\subseteq \{ \mbox{TSMF}, \mbox{MSMF}, \mbox{TUMF}, \mbox{MUMF} \}$ (black marks). For $th= 700$~G and $th=900$~G, the well manifested pink noise spectra inclined towards the red noise are seen, with $\left [ \left |m_{\Phi_{700}^{I}}\right | \right ]$ = [$1.2\pm0.15$, $1.8\pm0.15$] (blue marks), and $\left [ \left |m_{\Phi_{900}^{I}}\right | \right ]$ = [$1.4\pm0.15$, $1.9\pm0.15$] (red marks), $I\subseteq \{ \mbox{TSMF}, \mbox{MSMF}, \mbox{TUMF}, \mbox{MUMF} \}$, respectively.

\subsection{AR 12378}\label{AR_12378} 
This is the AR with the magnetic field dispersed in many small sized concentration areas, where the positive polarity areas are dominantly leading and the negative polarity ones are trailing.
The following features of the AR 12378 fluctuation noise spectra can be distinguished.

\begin{figure*}[ht!]
\center{\includegraphics[width=0.8\textwidth]{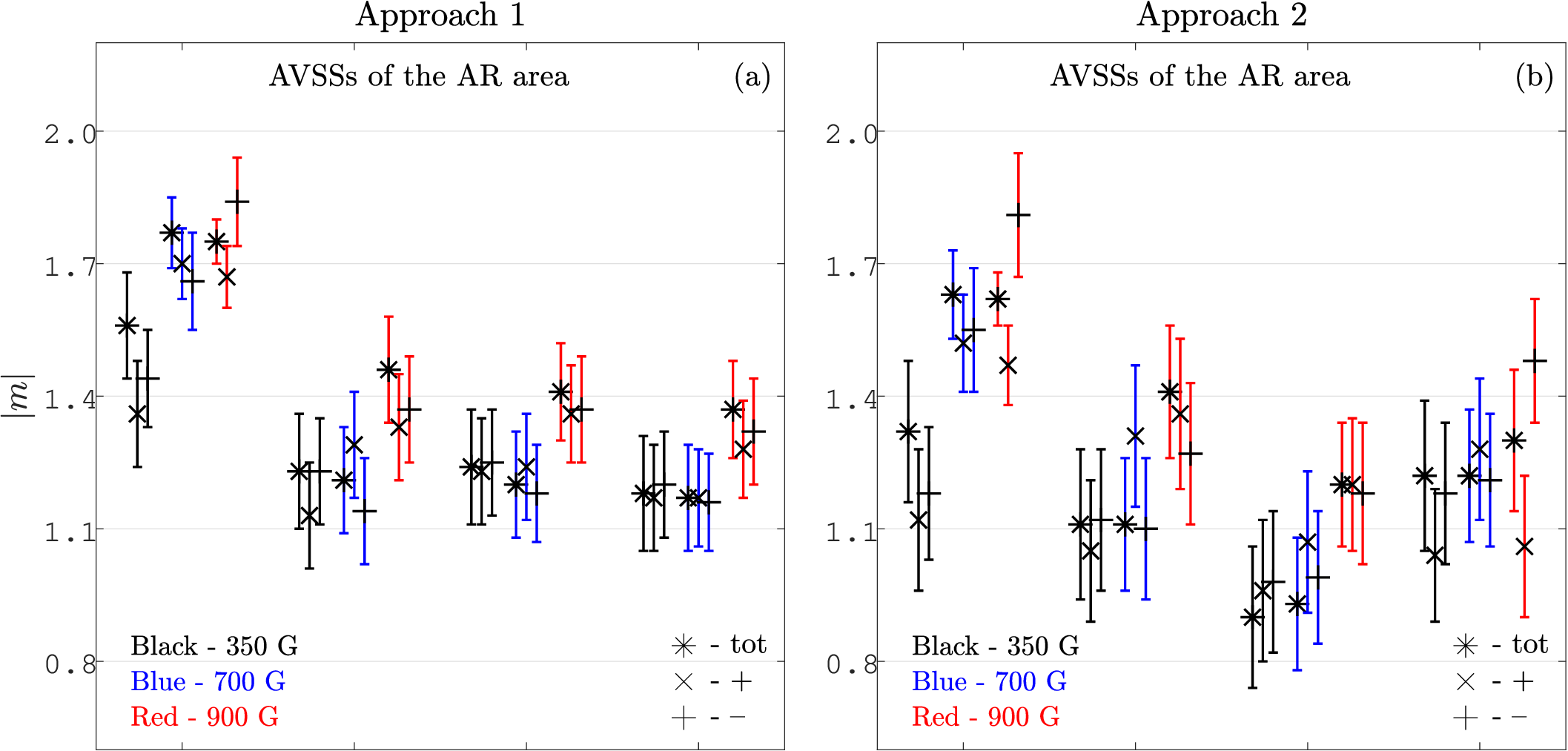}}
\center{\includegraphics[width=0.8\textwidth]{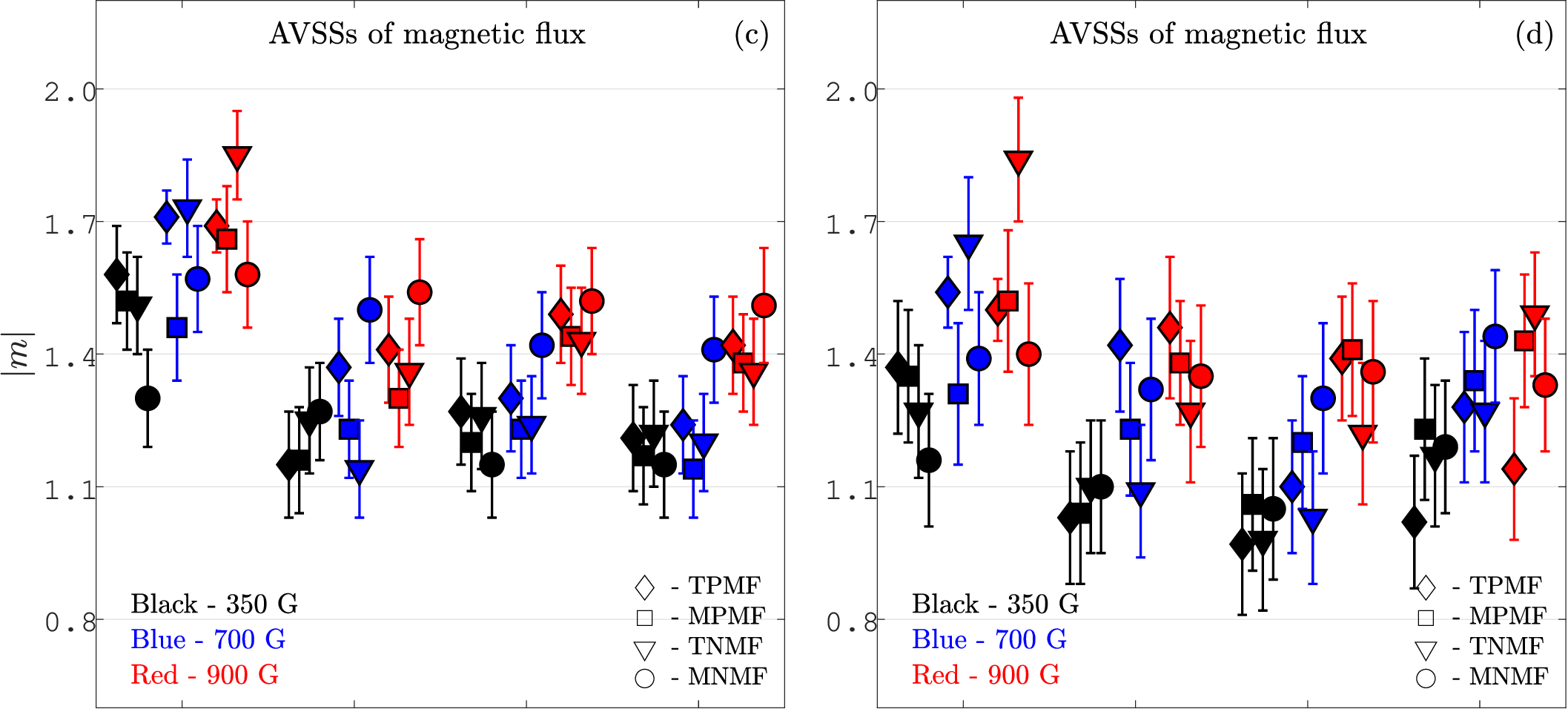}}
\center{\includegraphics[width=0.8\textwidth]{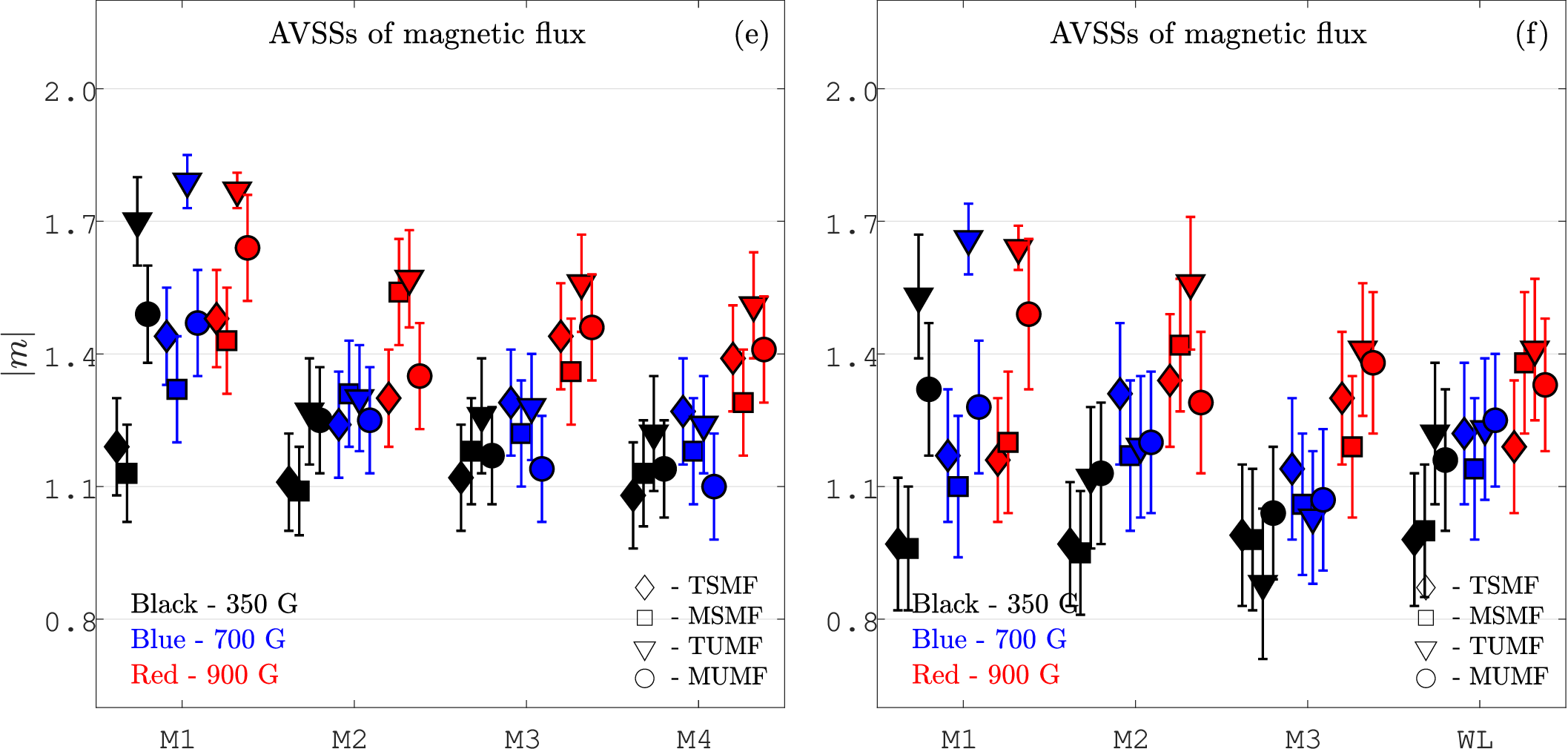}}
\caption{
Same as Fig.~\ref{results_12381}, but for the dispersed AR 12378.}
\label{results_12378}
\end{figure*}

(i) Contrary to the case of AR 12381, the AVSSs of the AR 12378 area fluctuations for the positive, negative, and total magnetic field, $\left |m _{\Sigma_{th}^{J}}\right |$, $J\subseteq \{ +, -, tot\}$, exhibit an arbitrary behaviour, without any systematic relation to each other (Fig.~\ref{results_12378} , a -b, shown with crosses, pluses, and asterisks, respectively). 

(ii) The spectral analysis in approach 1 (Fig.~\ref{results_12378}, a) shows that the AVSSs for the full set of AR areas, $\left |m _{\Sigma_{th}^{J}}\right |$, $th\subseteq \{ 350,700,900\}$, $J\subseteq \{ +, -, tot\}$, fluctuation noise range within [$1.1\pm0.12$, $1.5\pm0.12$], while  approach 2 (Fig.~\ref{results_12378}, b) yields the AVSSs within [$0.9\pm0.16$, $1.5\pm0.14$]. Contrary to the compact AR 12381, the AVSSs for the area fluctuations of the AR 12378 do not show any systematic monotonic dependence on the applied magnetic field threshold $th$, but they are rather arbitrarily distributed (see Table~\ref{table1} and \ref{table2} rows 1, 2, and 3). However, for $th= 700$ G, certain systematic interrelation is seen between the AVSSs of the area fluctuations obtained in the cases of the positive $\left |m _{\Sigma_{700}^{+}}\right |$ (Fig.~\ref{results_12378}, a-b, blue crosses) and negative $\left |m _{\Sigma_{700}^{-}}\right |$ (Fig.~\ref{results_12378}, a-b, blue pluses) magnetic field, specifically, the first ones are always larger than the latter. Moreover, for $th= 350$~G and $th= 700$~G, the spectra are close to the $1/f-$ type ones, with ranging within $\left [ \left |m _{\Sigma_{350}^{J}}\right | \right ]$ = [$0.9\pm0.16$, $1.3\pm0.12$] (black marks) and $\left [ \left |m _{\Sigma_{700}^{J}}\right | \right ]$ = [$0.9\pm0.15$, $1.3\pm0.16$] (blue marks), $J\subseteq \{ +, -, tot\}$, respectively. Application of the threshold $th= 900$ G yields a well manifested pink noise spectrum with the AVSSs of subset ranging within $\left [ \left |m _{\Sigma_{900}^{J}}\right | \right ]$ = [$1.1\pm0.16$, $1.5\pm0.14$], $J\subseteq \{ +, -, tot\}$ (red marks). 

(iii) Unlike the case of AR 12381, the AVSSs for the positive (leading sunspot group) and negative (trailing sunspot group) total, namely, $\left |m_{\Phi_{th}^{I}}\right |$, $I\subseteq \{ \mbox{TPMF}, \mbox{TNMF} \}$ (diamonds and triangles), and mean, namely, $\left |m_{\Phi_{th}^{I}}\right |$, $I\subseteq \{ \mbox{MPMF}, \mbox{MNMF} \}$ (squares and circles) magnetic fluxes in this AR do not show any monotonic interrelation, but are rather arbitrarily distributed in various constructed spectra (Fig.~\ref{results_12378}, c-d). 

(iv) Similarly to the case of different polarities, the AVSSs of both the signed and unsigned total, namely, $\left |m_{\Phi_{th}^{I}}\right |$, $I\subseteq \{ \mbox{TSMF}, \mbox{TUMF} \}$ (diamonds and triangles) and mean, that is, $\left |m _{\Phi_{th}^{I}}\right |$, $I\subseteq \{ \mbox{MSMF}, \mbox{MUMF} \}$ (squares and circles) also have irregularly distributed values in the spectra (Fig.~\ref{results_12378}, e-f). 

(v) The spectral analysis by approach 1 (Fig.~\ref{results_12378}, c) shows that the AVSSs of fluctuation noise for the subset of the ARs magnetic fluxes $\left |m_{\Phi_{th}^{I}}\right |$, $th\subseteq \{ 350,700,900\}$, $I\subseteq \{ \mbox{TPMF}, \mbox{MPMF}, \mbox{TNMF}, \mbox{MNMF} \}$, range within [$1.1\pm0.11$, $1.5\pm0.12$]; while  approach 2 (Fig.~\ref{results_12378}, d) yields the AVSSs within [$1.0\pm0.16$, $1.5\pm0.14$]. 
The AVSSs for the fluctuation spectra of magnetic flux of both polarities somewhat increase with the applied magnetic field threshold $th$ in this AR (see Table~\ref{table1} and \ref{table2} rows 4, 5, and 6). However, the ranges of their values are significantly narrower than in the case of AR 12381. Moreover, the spectra for all applied thresholds are close to the $1/f-$ type one, with the AVSSs for unified subset for both approaches of ranging within $\left [ \left |m_{\Phi_{350}^{I}}\right | \right ]$ = [$1.0\pm0.16$, $1.3\pm0.12$] (black marks), $\left [ \left |m_{\Phi_{700}^{I}}\right | \right ]$ = [$1.0\pm0.15$, $1.5\pm0.12$] (blue marks), and $\left [ \left |m_{\Phi_{900}^{I}}\right | \right ]$ = [$1.1\pm0.16$, $1.5\pm0.14$] (red marks), $I\subseteq \{ \mbox{TPMF}, \mbox{MPMF}, \mbox{TNMF}, \mbox{MNMF} \}$, respectively. 

(vi) The situation is similar in the cases of signed and unsigned magnetic flux fluctuation spectra. In particular, approach 1 (Fig.~\ref{results_12378}, e) reveals the AVSSs for the fluctuation noise of the signed and unsigned magnetic flux $\left |m_{\Phi_{th}^{I}}\right |$, $th\subseteq \{ 350,700,900\}$, $I\subseteq \{ \mbox{TSMF}, \mbox{MSMF}, \mbox{TUMF}, \mbox{MUMF} \}$, fall in the range within [$1.1\pm0.12$, $1.6\pm0.11$]; while approach 2 (Fig.~\ref{results_12378}, f) has the AVSSs range of [$0.9\pm0.17$, $1.6\pm0.15$]. Again, the AVSSs for the signed and unsigned magnetic flux fluctuations somewhat increase with the applied magnetic field threshold, $th,$ (see Table~\ref{table1} and \ref{table2} rows 7, 8, and 9). However, the ranges of their values are significantly narrower than in case of AR 12381. Moreover, the spectra for all the thresholds are close to the $1/f-$ type one, with the AVSSs for unified subset of both approaches ranging within $\left [ \left |m_{\Phi_{350}^{I}}\right | \right ]$ = [$0.9\pm0.17$, $1.3\pm0.13$] (black marks), $\left [ \left |m_{\Phi_{700}^{I}}\right | \right ]$ = [$1.0\pm0.15$, $1.3\pm0.16$] (blue marks), and $\left [ \left |m_{\Phi_{900}^{I}}\right | \right ]$ = [$1.2\pm0.16$, $1.6\pm0.16$] (red marks), $I\subseteq \{ \mbox{TSMF}, \mbox{MSMF}, \mbox{TUMF}, \mbox{MUMF} \}$, respectively.

\subsection{AR 12524}\label{AR_12524} 
This is an AR with a mixed internal structure, having two relatively large spots with opposite polarities (similar to the ones constituting the compact AR 12381) surrounded by many small areas of magnetic field concentration predominantly of the same polarity that are similar to the case of the dispersed AR 12378. The negative polarity areas are dominantly leading and the positive polarity are the trailing ones. The following features of the AR 12524 fluctuation noise spectra can be distinguished.

\begin{figure*}[ht!]
\center{\includegraphics[width=0.8\textwidth]{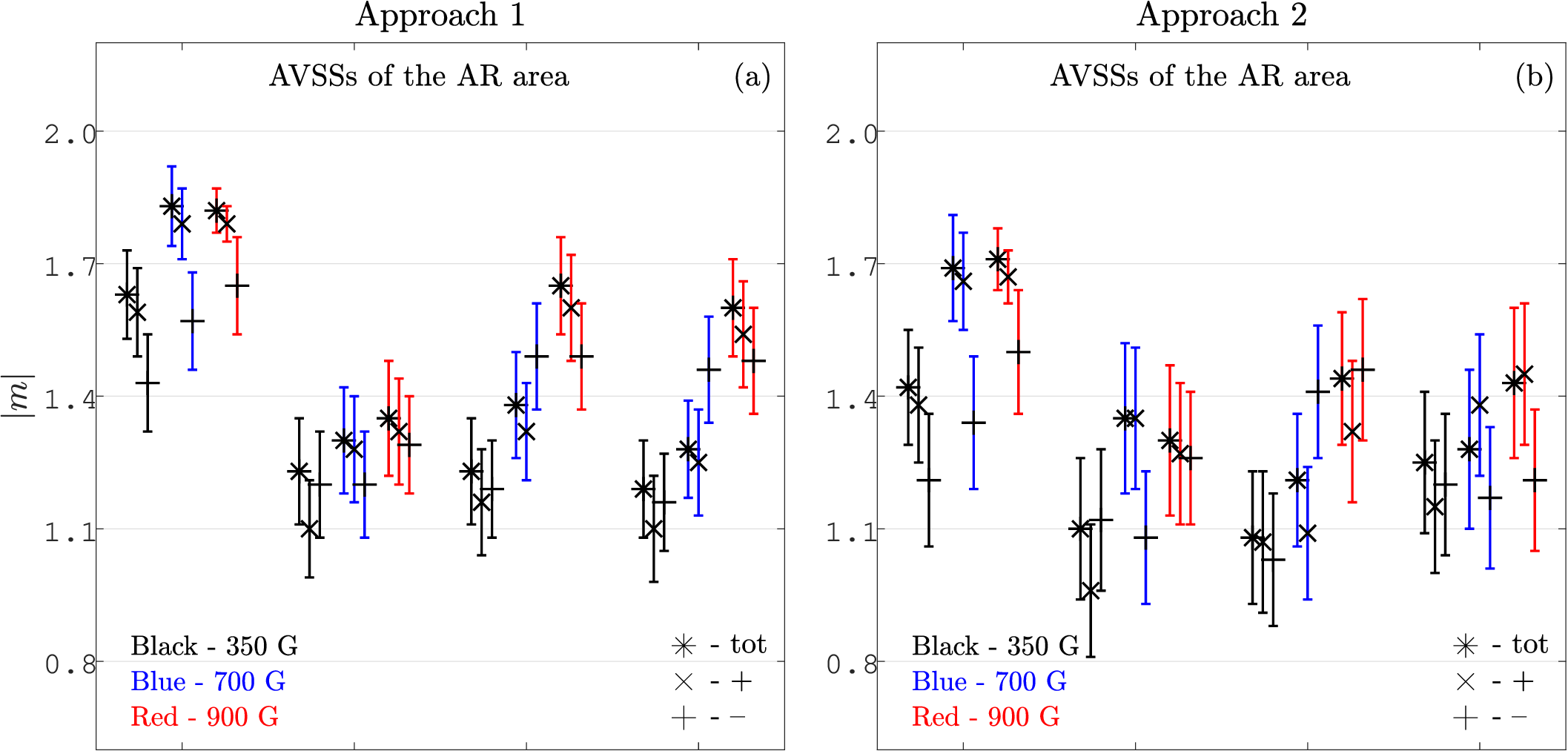}}
\center{\includegraphics[width=0.8\textwidth]{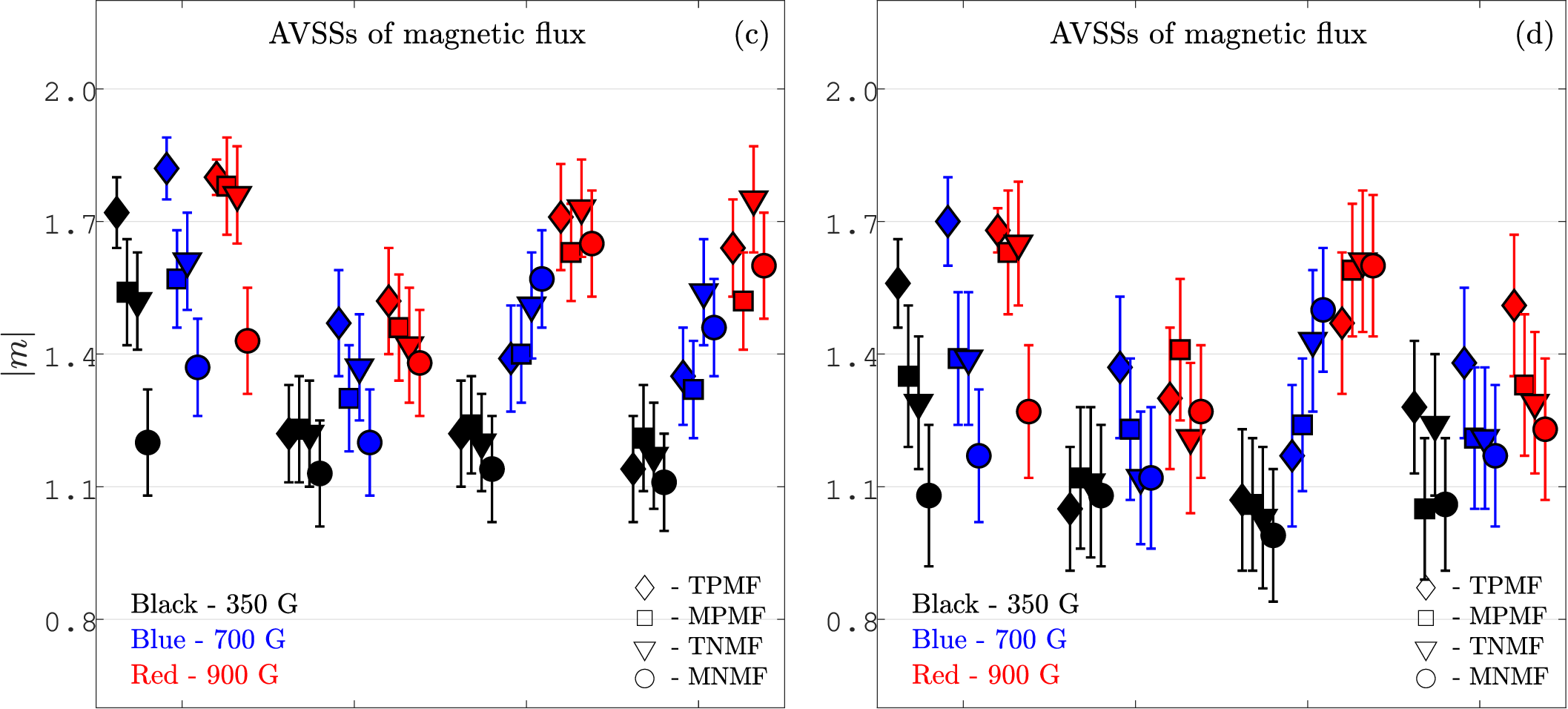}}
\center{\includegraphics[width=0.8\textwidth]{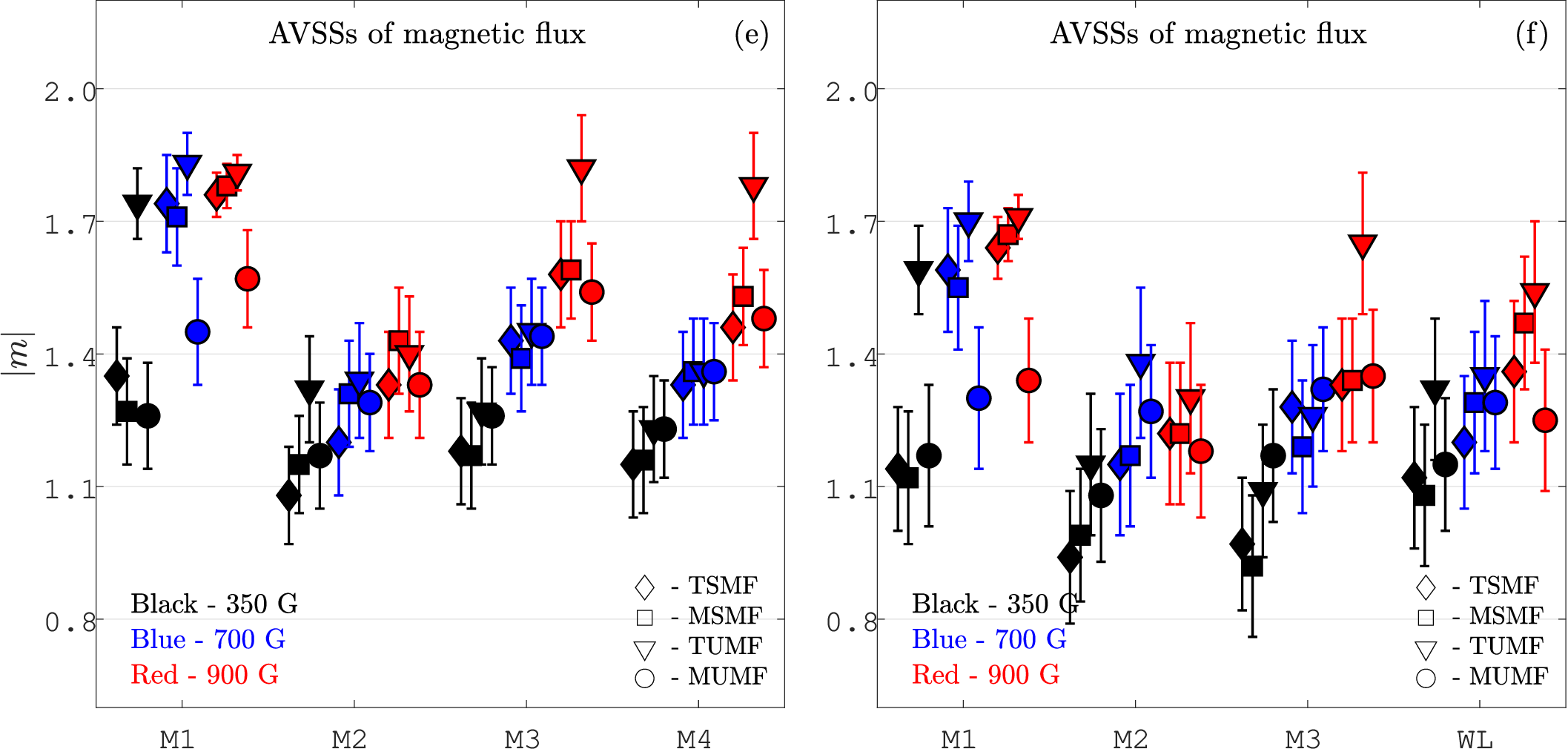}}
\caption{Same as Fig.~\ref{results_12381}, but for the mixed-type AR 12524.}
\label{results_12524}
\end{figure*}

(i) Again, contrary to the case of AR 12381 (and similarly to the case of AR 12378), the AVSSs of the AR 12524 area fluctuations for the positive, negative, and total magnetic field, $\left |m _{\Sigma_{th}^{J}}\right |$, $J\subseteq \{ +, -, tot\}$, exhibit arbitrary values that do not reveal any systematic relation between them (Fig.~\ref{results_12524}, a-b; shown as crosses, pluses, and asterisks, respectively). 

(ii) The spectral analysis via approach 1 (Fig.~\ref{results_12524}, a) shows that the AVSSs for the full set of AR areas, $\left |m _{\Sigma_{th}^{J}}\right |$, $th\subseteq \{ 350,700,900\}$, $J\subseteq \{ +, -, tot\}$, fluctuation noise range within [$1.1\pm0.11$, $1.6\pm0.11$], while approach 2 (Fig.~\ref{results_12524}, b) yields the AVSSs within [$1.0\pm0.15$, $1.5\pm0.16$]. The picture of the area fluctuation spectra in the case of this mixed type AR has partially the properties of both types of the ARs considered above; in particular, this refers to the AVSSs for the area, namely, the $\left |m _{\Sigma_{th}^{J}}\right |$, $th\subseteq \{ 350,700,900\}$, and $J\subseteq \{ +, -, tot\}  $  fluctuations of the AR 12524 (in some of the  spectrum measurement methods used here) show a certain monotonic increase, depending on the applied increasing magnetic field threshold, $th,$ (similar to the compact AR 12381), whereas this dependence is absent in the rest of the used methods (similar to the dispersed AR 12378, see Table~\ref{table1} and \ref{table2} rows 1, 2, and 3). At the same time, for $th= 350$~G and $th= 700$~G, the spectra are close to the $1/f-$ type ones with the AVSSs of subset ranging within $\left [ \left |m _{\Sigma_{350}^{J}}\right | \right ]$ = [$1.0\pm0.15$, $1.3\pm0.16$] (black marks) and $\left [ \left |m _{\Sigma_{700}^{J}}\right | \right ]$ = [$1.1\pm0.15$, $1.5\pm0.14$] (blue marks), respectively, while the well manifested pink noise spectrum with the AVSSs of subset within $\left [ \left |m _{\Sigma_{900}^{J}}\right | \right ]$ = [$1.2\pm0.16$, $1.6\pm0.11$] (red marks), $J\subseteq \{ +, -, tot\}$. 

(iii) In this AR, the AVSSs for the negative (leading sunspot group) and positive (trailing sunspot group) total, namely, $\left |m_{\Phi_{th}^{I}}\right |$, $I\subseteq \{ \mbox{TNMF}, \mbox{TPMF} \}$ (triangles and diamonds), and mean, namely, $\left |m_{\Phi_{th}^{I}}\right |$, $I\subseteq \{ \mbox{MNMF}, \mbox{MPMF} \}$ (circles  and squares) magnetic fluxes do not show any monotonic interrelation similar to the case of the dispersed AR 12378 (Fig.~\ref{results_12524}, c-d). 

(iv) Similarly to the case of different polarities, the AVSSs of both the signed and unsigned total, namely, $\left |m_{\Phi_{th}^{I}}\right |$, $I\subseteq \{ \mbox{TSMF}, \mbox{TUMF} \}$ (diamonds and triangles) and mean, i.e. $\left |m_{\Phi_{th}^{I}}\right |$, $I\subseteq \{ \mbox{MSMF}, \mbox{MUMF} \}$ (squares and circles) also have irregularly distributed values in the spectra (Fig.~\ref{results_12524}, e-f). 

(v) The spectral analysis by approach 1 (Fig.~\ref{results_12524}, c) shows that the AVSSs of fluctuation noise for the subset of the ARs magnetic fluxes, $\left |m_{\Phi_{th}^{I}}\right |$, $th\subseteq \{ 350,700,900\}$, $I\subseteq \{ \mbox{TPMF}, \mbox{MPMF}, \mbox{TNMF}, \mbox{MNMF} \}$, range within [$1.1\pm0.11$, $1.8\pm0.12$], while approach 2 (Fig.~\ref{results_12524} (d)) yields the AVSSs wihin [$1.0\pm0.15$, $1.6\pm0.16$]. The AVSSs for the fluctuation spectra of the magnetic flux of both polarities systematically increase with the applied magnetic field threshold $th$ as in the compact AR 12381 (see Table~\ref{table1} and \ref{table2} rows 4, 5, 6). Accordingly, for $th= 350$ G, the spectrum is close to the $1/f-$ type one, with the AVSSs for unified subset for both approaches of ranging within $\left [ \left |m_{\Phi_{350}^{I}}\right | \right ]$ = [$1.0\pm0.15$, $1.3\pm0.15$], $I\subseteq \{ \mbox{TPMF}, \mbox{MPMF}, \mbox{TNMF}, \mbox{MNMF} \}$ (black marks). For $th= 700$~G, the well manifested pink noise spectrum with the AVSSs within $\left [ \left |m_{\Phi_{700}^{I}}\right | \right ]$ = [$1.1\pm0.15$, $1.6\pm0.11$] (blue marks) is observed, which transforms for $th= 900$~G to the pink inclined towards the red noise spectrum with the AVSSs within $\left [ \left |m_{\Phi_{900}^{I}}\right | \right ]$ = [$1.2\pm0.17$, $1.8\pm0.12$] (red marks), $I\subseteq \{ \mbox{TPMF}, \mbox{MPMF}, \mbox{TNMF}, \mbox{MNMF} \}$. 

(vi) The situation is similar in the cases of the signed and unsigned magnetic flux fluctuation spectra. In particular, approach 1 (Fig.~\ref{results_12524} (e)) reveals the AVSSs for the fluctuation noise of the signed and unsigned magnetic flux $\left |m_{\Phi_{th}^{I}}\right |$, $th\subseteq \{ 350,700,900\}$, $I\subseteq \{ \mbox{TSMF}, \mbox{MSMF}, \mbox{TUMF}, \mbox{MUMF} \}$, fall in the range within [$1.1\pm0.11$, $1.8\pm0.12$], while approach 2 (Fig.~\ref{results_12524} (f)) gives the AVSSs within [$0.9\pm0.16$, $1.7\pm0.16$]. Again, the AVSSs for signed and unsigned magnetic flux fluctuations depend on the applied magnetic field threshold $th$ stronger than those in the case of the dispersed AR 12378 (see Table~\ref{table1} and \ref{table2} rows 7,8,9). 
Furthermore, for $th = 350$~G, the spectrum is close to the $1/f-$ type one with the AVSSs for unified subset of both approaches ranging within $\left [ \left |m_{\Phi_{350}^{I}}\right | \right ]$ = [$0.9\pm0.16$, $1.3\pm0.16$], $I\subseteq \{ \mbox{TSMF}, \mbox{MSMF}, \mbox{TUMF}, \mbox{MUMF} \}$ (black marks). For $th = 700$~G, the well manifested pink noise spectrum with the AVSSs within $\left [ \left |m_{\Phi_{700}^{I}}\right | \right ]$ = [$1.2\pm0.16$, $1.5\pm0.12$] (blue marks) is observed, which again transforms for $th = 900$~G to the pink inclined towards the red noise spectrum with the AVSSs within $\left [ \left |m_{\Phi_{900}^{I}}\right | \right ]$ = [$1.2\pm0.15$, $1.8\pm0.12$] (red marks), $I\subseteq \{ \mbox{TSMF}, \mbox{MSMF}, \mbox{TUMF}, \mbox{MUMF} \}$.

\subsection{AR 12435 and AR 12437}\label{AR_12435} 
These two unipolar active regions both have a dispersed internal structure (similar to AR 12378). The AR 12435 has a positive polarity and appears in a leading position in the considered sunspot group, whereas the AR 12437 has a negative polarity and is in the trailing position. The following features of the fluctuation noise spectra in the complex of AR 12435 and AR 12437 can be distinguished.

\begin{figure*}[ht!]
\center{\includegraphics[width=0.8\textwidth]{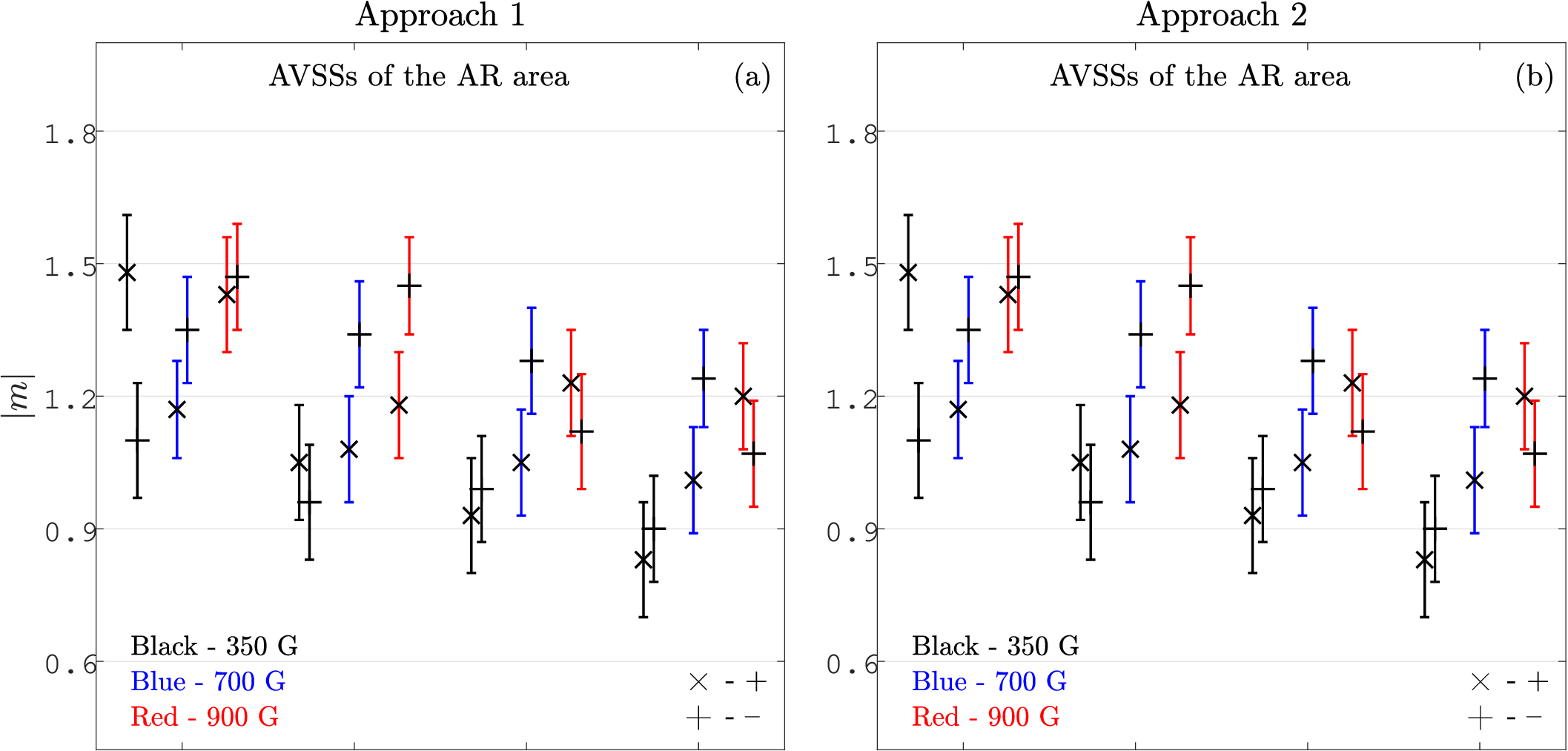}}
\center{\includegraphics[width=0.8\textwidth]{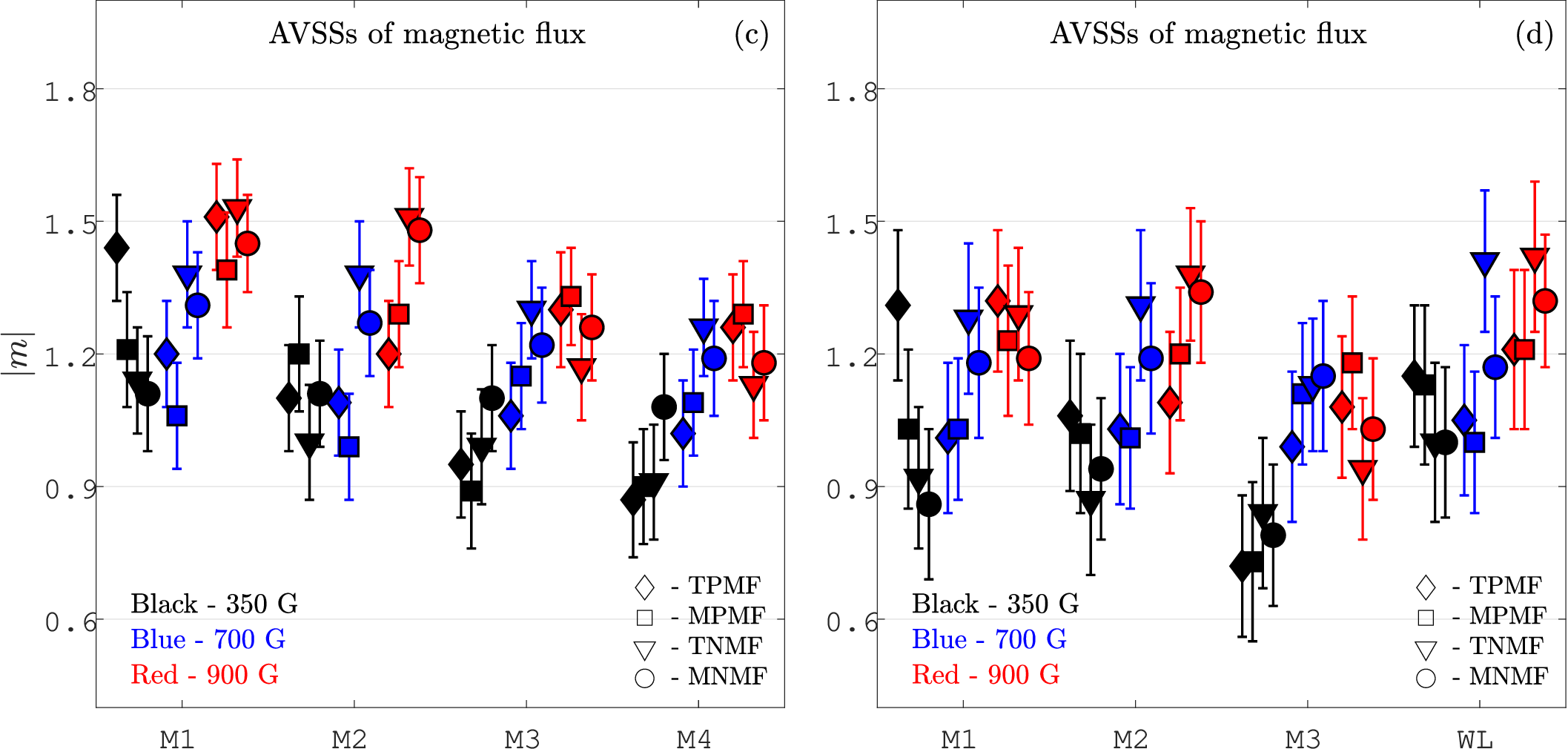}}
\caption{Same  as panels (a), (b), (c), and (d) in Fig.~\ref{results_12381}, but for the set of two unipolar ARs, AR 12435 and AR 12437, of different polarity. In panels (a) and (b), the markers '$\times$' and '$+$' denote the $\left |m _{\Sigma_{th}^{+,-}}\right |$ of the AR 12435 and AR 12437, respectively. In panels (c) and (d), the markers '$\diamond$' and '$\square$' indicate the $\left |m _{\Phi_{th}^{I}}\right |$, $I\subseteq \{ \mbox{TPMF}, \mbox{MPMF} \}$ for the AR 12435, whereas '$\triangle$' and '$\circ$' signify the $\left |m _{\Phi_{th}^{I}}\right |$, $I\subseteq \{ \mbox{TNMF}, \mbox{MNMF} \}$ for the AR 12437.}
\label{results_12435_12437}
\end{figure*}

(i) The AVSSs of the areas, $\left |m _{\Sigma_{th}^{+,-}}\right |$, fluctuations of these two ARs interrelate to each other similarly to those of the dispersed AR 12378 (Fig.~\ref{results_12435_12437}, a-b), with crosses and pluses for AR 12435 and AR 12437, respectively.

(ii) The spectral analysis via approach 1 (Fig.~\ref{results_12435_12437}, a) shows that the AVSSs for the full set of AR areas, fluctuation noise for AR 12435, $\left |m _{\Sigma_{th}^{+}}\right |$, $th\subseteq \{ 350,700,900\}$, and AR 12437, $\left |m _{\Sigma_{th}^{-}}\right |$, $th\subseteq \{ 350,700,900\}$, range within [$0.8\pm0.13$, $1.2\pm0.12$] and [$0.9\pm0.12$, $1.5\pm0.11$], respectively. Approach 2 (Fig.~\ref{results_12435_12437}, b) yields the AVSSs within [$0.7\pm0.17$, $1.2\pm0.17$] and [$0.8\pm0.17$, $1.4\pm0.18$] for the AR 12435 and AR 12437, respectively. Similarly to the dispersed AR 12378, the AVSSs for the area $\left |m _{\Sigma_{th}^{+,-}}\right |$, fluctuations of both considered unipolar ARs do not show any systematic dependence on the applied magnetic field thresholds $th$ (see Table~\ref{table1} and \ref{table2} rows 1, 2, and 3). The only systematic interrelation seen, is that the AVSSs for the area $\left |m _{\Sigma_{th}^{-}}\right |$, fluctuations of AR 12437 (blue pluses) are always higher than those of $\left |m _{\Sigma_{th}^{+}}\right |$ of the AR 12435 (blue crosses). 
Furthermore, for all applied magnetic field thresholds $th$, the spectra are close to $1/f-$ type for both ARs with the AVSSs ranging within: 
\begin{align*}
&\left [ \left |m _{\Sigma_{350}^{+}}\right | \right ] = [0.7\pm0.17, 1.1\pm0.17]~ \text{and}~\\
&\left [ \left |m _{\Sigma_{350}^{-}}\right | \right ] = [0.8\pm0.17, 1.0\pm0.12]~\text{(black marks);} \\
\\[0.1cm] 
&\left [ \left |m _{\Sigma_{700}^{+}}\right | \right ] = [1.0\pm0.16, 1.1\pm0.16]~ \text{and}~\\
&\left [ \left |m _{\Sigma_{700}^{-}}\right | \right ] = [1.1\pm0.15, 1.4\pm0.17] ~\text{(blue marks);} \\
\\[0.1cm]
&\left [ \left |m _{\Sigma_{900}^{+}}\right | \right ] = [1.0\pm0.16, 1.2\pm0.17]~ \text{and}~\\
&\left [ \left |m _{\Sigma_{700}^{-}}\right | \right ] = [0.9\pm0.16, 1.5\pm0.11] ~\text{(red marks)} 
\end{align*}
for the AR 12435 and AR 12437, respectively.

(iii) Similarly to the dispersed AR 12378, the AVSSs for the positive (AR 12435) and negative (AR 12437) total, namely, $\left |m_{\Phi_{th}^{I}}\right |$, $I\subseteq \{ \mbox{TPMF}, \mbox{TNMF} \}$ (diamonds and triangles), and mean, i.e. $ \left |m_{\Phi_{th}^{I}}\right |$, $I\subseteq \{ \mbox{MPMF}, \mbox{MNMF} \}$ (squares and circles) magnetic flux do not show any monotonic interrelation, but are rather arbitrarily distributed in various constructed spectra (Fig.~\ref{results_12435_12437}, c-d). 

(iv) The analysis of the signed and unsigned magnetic fluxes is irrelevant in this case of two unipolar ARs, whereas the case of total and mean magnetic fluxes is addressed above in point (iii). 

(v) The spectral analysis by approach 1 (Fig.~\ref{results_12435_12437}, c) shows that the AVSSs of fluctuation noise for the subset of the ARs magnetic fluxes $\left |m_{\Phi_{th}^{I}}\right |$, $th\subseteq \{ 350,700,900\}$, range within [$0.9\pm0.13$, $1.3\pm0.11$] and [$0.9\pm0.13$, $1.5\pm0.11$], for the AR 12435 ($I\subseteq \{ \mbox{TPMF}, \mbox{MPMF} \}$) and AR 12437 ($I\subseteq \{ \mbox{TNMF}, \mbox{MNMF} \}$), respectively. 
Accordingly, approach 2 (Fig.~\ref{results_12435_12437} (d)) yields the AVSSs within [$0.7\pm0.16$, $1.2\pm0.18$] and [$0.8\pm0.16$, $1.4\pm0.17$], for the AR 12435 and AR 12437, respectively. As in the case of the dispersed AR 12378, the AVSSs for the fluctuation spectra of magnetic flux of both ARs slightly increase with the applied magnetic field thresholds, $th$, and their ranges are narrower, as compared to those of the compact AR 12381 (see Table~\ref{table1} and \ref{table2} rows 4, 5, and 6). 
Furthermore, the spectra obtained for all $th$ are close to the $1/f-$ type spectrum, which is evidenced by the corresponding AVSSs ranges:
\begin{align*}
&\left [ \left |m_{\Phi_{350}^{TPMF,MPMF}}\right | \right ] = [0.7\pm0.16, 1.2\pm0.16]~ \text{and}~\\
&\left [ \left |m_{\Phi_{350}^{TNMF,MNMF}}\right | \right ] = [0.8\pm0.16, 1.1\pm0.12] ~\text{(black marks);} \\ 
\\[0.1cm]
&\left [ \left |m_{\Phi_{700}^{TPMF,MPMF}}\right | \right ] = [1.0\pm0.17, 1.2\pm0.12]~ \text{and}~\\
&\left [ \left |m_{\Phi_{700}^{TNMF,MNMF}}\right | \right ] = [1.1\pm0.15, 1.4\pm0.16] ~\text{(blue marks);} \\
\\[0.1cm]
&\left [ \left |m_{\Phi_{900}^{TPMF,MPMF}}\right | \right ] = [1.1\pm0.16, 1.3\pm0.11]~ \text{and}~\\
&\left [ \left |m_{\Phi_{900}^{TNMF,MNMF}}\right | \right ] = [0.9\pm0.16, 1.5\pm0.11] ~\text{(red marks);} 
\end{align*}
for the AR 12435 and AR 12437, respectively.

(vi) The comparison of approaches 1 and 2 for the analysis of the signed and unsigned magnetic fluxes is irrelevant in this case of two unipolar ARs.


\end{appendix}

\end{document}